\documentclass[twocolumn,twocolappendix]{aastex63}
\pdfoutput=1
\usepackage{amsmath,amstext}
\usepackage[T1]{fontenc}
\usepackage{times}
\usepackage{apjfonts} 
\usepackage[figure,figure*]{hypcap}
\usepackage{natbib}
\usepackage{hyperref}
\usepackage{booktabs}
\usepackage{graphicx}
\usepackage{multirow}
\usepackage[utf8x]{inputenc} 
\usepackage[caption=false]{subfig}

\received{2021 August 18}
\revised{2021 October 18}
\accepted{2021 October 26}
\submitjournal{the Astrophysical Journal}
\shorttitle{Diagnosing DASH}
\shortauthors{Cutler et al.}

\begin{document}
\title{Diagnosing DASH: A Catalog of Structural Properties for the COSMOS-DASH Survey}

\correspondingauthor{Sam E. Cutler}
\email{secutler@umass.edu}

\author[0000-0002-7031-2865]{Sam E. Cutler}
\affiliation{Department of Astronomy, University of Massachusetts, Amherst, MA 01003, USA}
\author[0000-0001-7160-3632]{Katherine E. Whitaker}
\affiliation{Department of Astronomy, University of Massachusetts, Amherst, MA 01003, USA}
\affiliation{Cosmic Dawn Center (DAWN), Copenhagen, Denmark}
\author[0000-0002-8530-9765]{Lamiya A. Mowla}
\affiliation{Dunlap Institute for Astronomy and Astrophysics, University of Toronto, 50 St. George Street, Toronto, ON M5S 3H4, Canada}
\author[0000-0003-2680-005X]{Gabriel B. Brammer}
\affiliation{Cosmic Dawn Center (DAWN), Copenhagen, Denmark}
\affiliation{Niels Bohr Institute, University of Copenhagen, Lyngbyvej 2, 2100 Copenhagen, Denmark}
\author[0000-0002-5027-0135]{Arjen van der Wel}
\affiliation{Sterrenkundig Observatorium, Universiteit Gent, Krijgslaan 281 S9, B-9000 Gent, Belgium}
\author[0000-0001-9002-3502]{Danilo Marchesini}
\affiliation{Department of Physics and Astronomy, Tufts University, Medford, MA 02153, USA}
\author[0000-0002-8282-9888]{Pieter G. van Dokkum}
\affiliation{Astronomy Department, Yale University, New Haven, CT 06511, USA}
\author[0000-0003-1665-2073]{Ivelina G. Momcheva}
\affiliation{Space Telescope Science Institute, 3700 San Martin Dr., Baltimore, MD 21211, USA}
\author[0000-0002-8442-3128]{Mimi Song}
\affiliation{Department of Astronomy, University of Massachusetts, Amherst, MA 01003, USA}
\author[0000-0002-3240-7660]{Mohammad Akhshik}
\affiliation{Department of Physics, University of Connecticut, Storrs, CT 06269, USA}
\author[0000-0002-7524-374X]{Erica J. Nelson}
\affiliation{Department for Astrophysical and Planetary Science, University of Colorado, Boulder, CO 80309, USA}
\author[0000-0001-5063-8254]{Rachel Bezanson}
\affiliation{University of Pittsburgh, Department of Physics and Astronomy, 100 Allen Hall, 3941 O'Hara St., Pittsburgh, PA 15260, USA}
\author[0000-0002-8871-3026]{Marijn Franx}
\affiliation{Leiden Observatory, Leiden University, P.O. Box 9513, NL-2300 AA Leiden, The Netherlands}
\author[0000-0002-7613-9872]{Mariska Kriek}
\affiliation{Astronomy Department, University of California, Berkeley, CA 94720, USA}
\author[0000-0001-6755-1315]{Joel Leja}
\affiliation{Department of Astronomy \& Astrophysics, The Pennsylvania State University, University Park, PA 16802, USA}
\affiliation{Institute for Computational \& Data Sciences, The Pennsylvania State University, University Park, PA 16802, USA}
\affiliation{Institute for Gravitation and the Cosmos, The Pennsylvania State University, University Park, PA 16802, USA}
\author[0000-0001-6529-8416]{John W. MacKenty}
\affiliation{Space Telescope Science Institute, 3700 San Martin Dr., Baltimore, MD 21211, USA}
\author[0000-0002-9330-9108]{Adam Muzzin}
\affiliation{Department of Physics and Astronomy, York University 4700 Keele St., Toronto, Ontario, Canada MJ3 1P3}
\author{Heath Shipley}
\affiliation{Department of Physics, McGill University, 3600 Rue University, Montr\'eal, QC H3P 1T3, Canada}
\author{Daniel Lange-Vagle}
\affiliation{Department of Physics and Astronomy, Tufts University, Medford, MA 02153, USA}

\begin{abstract}
We present the $H_{\mathrm{160}}$ morphological catalogs for the COSMOS-DASH survey, the largest area near-IR survey using HST-WFC3 to date. Utilizing the ``Drift And SHift'' observing technique for HST-WFC3 imaging, the COSMOS-DASH survey imaged approximately 0.5 deg$^2$ of the UltraVISTA deep stripes (0.7 deg$^2$ when combined with archival data). Global structural parameters are measured for 51,586 galaxies within COSMOS-DASH using GALFIT (excluding the CANDELS area) with detection using a deep multi-band HST image. We recover consistent results with those from the deeper 3D-HST morphological catalogs, finding that, in general, sizes and S\'ersic indices of typical galaxies are accurate to limiting magnitudes of $H_{160}<23$ and $H_{160}<22$ ABmag, respectively. In size-mass parameter space, galaxies in COSMOS-DASH demonstrate robust morphological measurements out to $z\sim2$ and down to $\log(M_\star/M_\odot)\sim9$. With the advantage of the larger area of COSMOS-DASH, we measure a flattening of the quiescent size-mass relation below $\log(M_\star/M_\odot)\sim10.5$ that persists out to $z\sim2$.  We show that environment is not the primary driver of this flattening, at least out to $z=1.2$, whereas internal physical processes may instead govern the structural evolution.
\end{abstract}


\section{Introduction}
Balancing resolution, depth, and area of optical/near-infrared (NIR) observations is critical in sampling the properties of galaxies across a wide range of redshifts and masses. From the ground, many surveys have been able to cover areas in the hundreds or thousands of square degrees \citep[e.g., the 5000 deg$^2$ optical Dark Energy Survey and the deep two deg$^2$ NIR UltraVISTA survey;][]{DES05,McCracken12} and have been important in advancing the study of galaxy formation and evolution. For example, UltraVISTA (UVISTA) has contributed to our understanding of massive galaxy structure, color, and mass out to $z\sim5$ \citep{Hill17}, as well as the evolution of the stellar mass function out to $z\sim4$ \citep{Muzzin13b}. Despite the many achievements of ground based surveys, the spatial resolution of these observations is limited by seeing, so deblending becomes important at higher redshift \citep{Mowla19b,Marsan19}. For high spatial resolution, the \textit{Hubble Space Telescope} (HST) is excellent, but its narrow FOV is not ideal for wide area surveys. The largest area observed with HST is the 1.7 square degree Cosmic Evolution Survey (COSMOS) field imaged using the optical F814W filter of the Advanced Camera for Surveys \citep[ACS;][]{Scoville07,Koekemoer07}. Until recently, the largest HST survey in the NIR by comparison is the 0.25 square degree Cosmic Assembly Near-infrared Deep Extragalactic Legacy Survey (CANDELS) survey in the F160W and F125W filters of the Wide Field Camera 3 (WFC3) \citep{Koekemoer11}.

The CANDELS survey in particular has led to a wealth of important results, many of which make full use of the HST-resolution morphological catalogs presented in \cite{vanderWel12} \citep[e.g.,][]{Bell12,Wuyts12,Barro13,vanderWel14,vanDokkum15,Barro17,Marsan19,Mowla19a,Mowla19b,Suess19a,Suess19b,Chen20}, as well as other morphological and structural measurements \citep[e.g.,][]{vanDokkum11,Kartaltepe12,Shibuya15,Holwerda15,Nelson16,Dimauro18,Dimauro19,Nedkova21}. To name a few, this data yields clear evidence for inside out disk growth \citep{Wuyts12,Nelson16}, establishes correlations between quiescence and galaxy morphology \citep{Bell12}, and uncovers the striking diversity of morphologies among massive galaxies at high redshift \citep{vanDokkum11}. Structural measurements from \cite{vanderWel12} are also fundamental to establishing significant correlations between the surface mass densities of star-forming and quiescent galaxies \citep[e.g.,][]{Whitaker17}; these results inspired astronomers to coin a new term, the structural main sequence, that hypothesizes a universal evolution track of compaction among star-forming galaxies \citep{Barro13,Barro17}. 

Combining procedures and imaging utilized in the CANDELS morphological catalogs by \cite{vanderWel12} with analysis from 3D-HST \citep{Brammer12,Skelton14,Momcheva16}, \cite{vanderWel14} perform the first comprehensive investigation of the size-mass distribution of galaxies out to $z=3$. \cite{vanderWel14} definitively establish that both quiescent and star forming galaxies of the same stellar mass are smaller in size at higher redshift, with the number density of compact, quiescent galaxies peaking around $z\sim1.5-2$ \citep[see also][]{Cassata11}. Building a sample of ultra-massive galaxies with HST-resolution morphologies, \cite{Mowla19b} extend the study of the evolution of the size-mass relation to $\log(M_\star/M_\odot)>11.3$ and $z=3$ \citep[see also][]{Marsan19,Nedkova21}. Contrary to scaling relations measured for less massive galaxies \citep[e.g.,][]{vanderWel14}, the most massive quiescent galaxies instead have similar sizes to star-forming galaxies of similar stellar mass at all redshifts explored \citep{Mowla19b}. Moreover, by combining the COSMOS-DASH and 3D-HST surveys, \cite{Mowla19a} show that the galaxy size-mass relation is best represented by a broken power law. These studies obtain comprehensive samples of galaxies owing to the combination of the deep, CANDELS-depth observations with the wide area observations of optical ($I_{814}$) COSMOS ACS \citep{Scoville07,Koekemoer07} and NIR ($H_{160}$) COSMOS-DASH imaging, the latter of which is possible due to the ``Drift And SHift'' (DASH) imaging technique \citep{Momcheva17}. 

While \cite{Mowla19b} analyzed only the most massive galaxies in COSMOS-DASH, the motivation for this work is to augment earlier work by releasing a full COSMOS-DASH morphological catalog.  The focus here will be on testing and quantifying the stellar mass and redshift limits of this moderately shallow imaging. In the following sub-sections, we first intro the DASH technique (Section \ref{sec:DASH}), followed by the COSMOS-DASH survey itself in more detail (Section \ref{sec:surv}).
\begin{figure*}
    \centering
    \includegraphics[width=\linewidth]{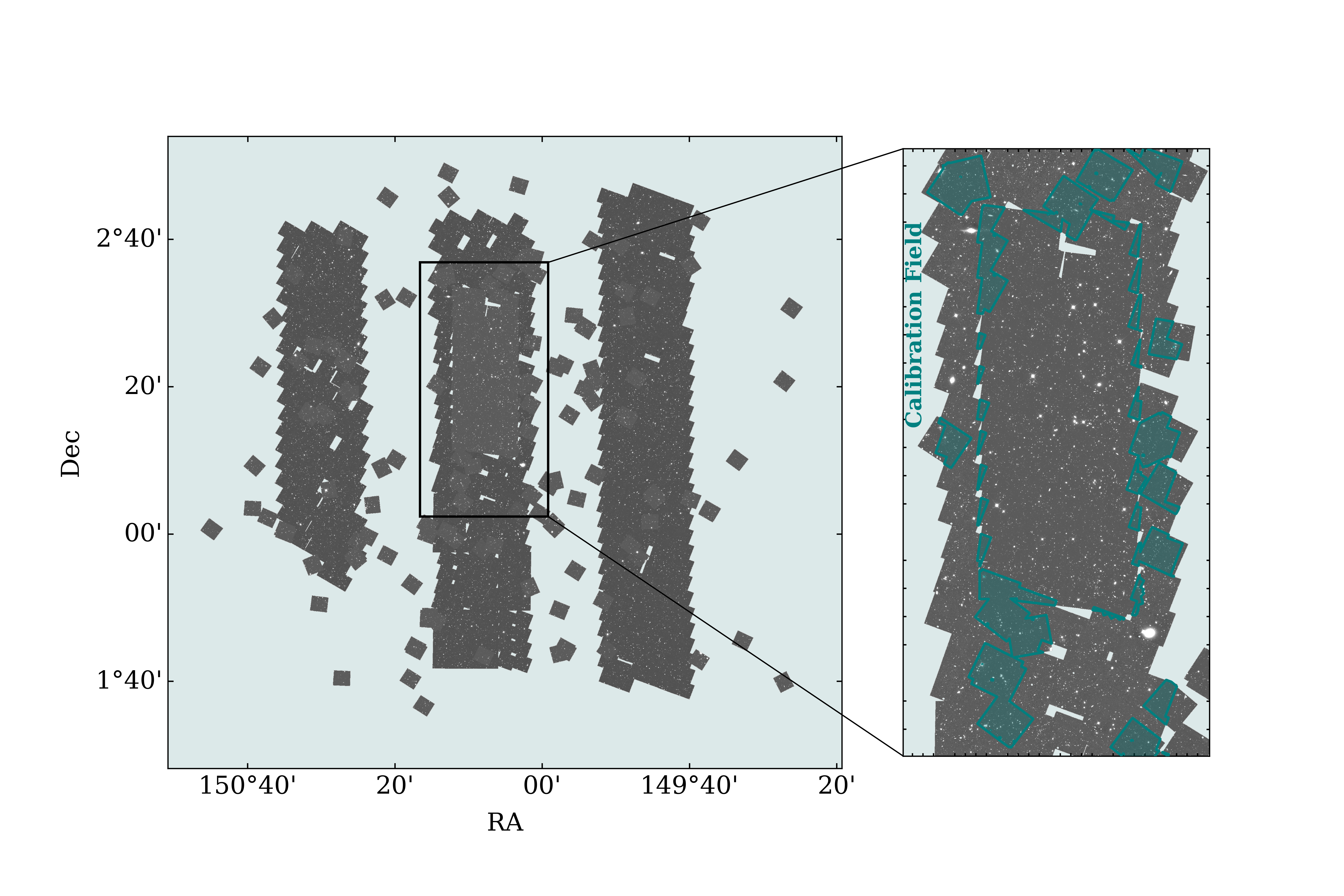}
    \caption{Image of the COSMOS-DASH mosaic, which roughly covers the UVISTA deep stripes. The left panel shows the science image of the COSMOS-DASH field with CANDELS and other archival exposures included. The inset shows the ``calibration field'', which is the region of COSMOS-DASH that overlaps with CANDELS. The overlap itself is highlighted in teal, and facilitates a direct comparison of DASH and CANDELS depth morphologies.}
    \label{fig:mosaic}
\end{figure*}

\subsection{Drift And SHift (DASH) Technique}
\label{sec:DASH}
Surveying large areas with HST is observationally expensive for two reasons. The first is the small, 4.6 arcmin$^2$ field-of-view (FOV), requiring roughly 750 pointings to map degree scale areas on the sky. However, just as significant, especially in the NIR, are the large overheads stemming from guide star acquisition. Standard HST observations require guide star acquisitions that take over 300s, often limiting observations to a single pointing per orbit. Hence, in the case of one orbit per pointing, covering degree-scale areas would require approximately 25\% of the annual pool available to the community, assuming 3000 orbits per HST cycle. The novel DASH technique \citep{Momcheva17} turns off guide star acquisition between pointings to allow up to eight, 300s pointings to be observed in a single orbit. In DASH, a single guide star is acquired for the first pointing after which the fine guidance sensors (FGS) are disabled and the gyros are used to slew to subsequent pointings. Without the use of the FGS, gyro bias introduces an instrument drift of 0.001\arcsec{}-0.002\arcsec{} per second. This drift smears exposures exceeding 60s, rendering them unusable. However, exposures from the WFC3 IR detector are composed of several zero-overhead reads, whose exposure times can be set to $<25$s, corresponding to a drift of $\lesssim0.05$\arcsec{}. A full 300s DASH exposure consists of 12 reads of 25s, which can be treated as 12 individual, dithered exposures. The independent reads are then shifted and drizzled in post-processing to restore the full HST resolution with high fidelity \citep[see Fig. 1 of][]{Momcheva17}. 

\subsection{COSMOS-DASH}\label{sec:surv}
Using the DASH technique, \cite{Mowla19b} present COSMOS-DASH, a 0.66 deg$^2$ NIR survey of the COSMOS field in the F160W filter of WFC3, 0.49 deg$^2$ of which are taken with 57 DASH orbits. With the first four COSMOS-DASH orbits, \cite{Momcheva17} demonstrate that HST resolution is recovered and the structural parameters are consistent with those from 3D-HST \citep{vanderWel12,vanderWel14} for objects  brighter than $H_{160}=22$ ABmag. \cite{Mowla19b} also fit the morphologies of 910 galaxies with $M_\star>2\times10^{11}M_\odot$ in COSMOS-DASH and COSMOS ACS \citep{Scoville07,Koekemoer07} and supplemented these with lower mass galaxies from the 3D-HST morphological catalogs \citep{vanderWel14} to measure the size-mass relation. 

While DASH allows for high resolution, wide-field imaging in fewer orbits, it does so at the cost of image depth. Since DASH exposures are limited to around 300s per pointing, the image depth is subsequently lower: the $5\sigma$ point source depth is 25.1 ABmag, compared to $\sim27$ ABmag for CANDELS \citep{Koekemoer11}. Moreover, shifting the individual reads in post-processing could also impact the ability to accurately measure morphologies of all but the most massive galaxies. The intermediate depth, combined with the unusual data reduction techniques suggest that DASH observations may be unsuitable for obtaining galaxy morphologies for fainter low-mass or extended galaxies. The exact parameter space in which morphological fits become unreliable is the study of this work. 

In this paper, we present the COSMOS-DASH morphological catalog, inclusive of sources in \cite{Mowla19b} but extending toward lower masses to yield a magnitude limited sample. Section \ref{sec:obs} describes the COSMOS-DASH survey and observations used to build this catalog. In Section \ref{sec:methods}, we describe methods employed to fit the morphological parameters and to determine the uncertainty on these parameters. Section \ref{sec:cat} discusses the format and layout of the morphological catalog itself. Our key results are presented in Section \ref{sec:res}, where we compare COSMOS-DASH morphologies to other samples and discuss the mass dependence of the size-mass relation and the roles galactic environment and internal physical processes play in it. A summary of results and future directions for this study are given in Section \ref{sec:sum}. We assume a flat $\Lambda$CDM cosmology with $\Omega_m=0.3$, $\Omega_\Lambda=0.7$, and $H_0=70\mathrm{~km~s}^{-1}~\mathrm{Mpc}^{-1}$, as well as a Chabrier IMF \citep{Chabrier03} for stellar masses. All magnitudes are in the AB system.

\section{Observations and Data}\label{sec:obs}

\subsection{COSMOS-DASH Mosaic}
The Cycle 23 COSMOS-DASH program (Program ID: GO-14114) is a 0.49 deg$^2$ survey of the COSMOS field using the HST-WFC3 F160W filter. It consists of 456 pointings in 57 orbits (8 pointings per orbit) obtained between 2016 November and 2017 June using the aforementioned DASH technique. Each exposure consists of 11 or 12 25s reads which are shifted and drizzled to remove the effect of instrument drift to thereby restore HST resolution \citep[see][]{Momcheva17}. The data set additionally includes all available archival F160W data in COSMOS, including CANDELS, resulting in a total F160W area of 0.66 deg$^{-2}$. We use an updated, redrizzled mosaic (containing more archival data) that has a pixel scale of 0.1\arcsec{} pix$^{-1}$, centered at R.A.=10:00:20.2, decl.=+2:11:03.7, shown in Fig. \ref{fig:mosaic}, main panel. 

\subsection{COSMOS-DASH Sample}\label{sec:sample}
For this project, we select galaxies from an HST-selected photometric catalog of sources in COSMOS-DASH. Sources are detected by running SExtractor v2.19.5 \citep{Bertin96} on a combined WFC3 (F105W, F110W, F125W, F140W, and F160W) and ACS (F814W) noise-equalized HST image, where available. We identify galaxies following the standard settings of e.g., \cite{Skelton14} and \cite{Shipley18}. The catalog aperture is 0.7\arcsec{} in diameter and all catalog sources are required to have coverage in at least one WFC3 band; sources with F814W coverage only are therefore not included.  The catalog is generated from a global, deep detection segmentation map that is used to mask objects for GALFIT (see Sec. \ref{sec:seg}). The primary purpose of this catalog herein is for source detection to facilitate the goal of measuring robust structural parameters and deblending lower-resolution ground based data. We therefore focus on the F160W photometry alone. Kron radii from SExtractor are also used in determining image cutout size (see Sec. \ref{sec:galfit}).

Cross-matching this catalog with the UltraVISTA catalog of \cite{Muzzin13a}, we identify 51,586 galaxies covered by COSMOS-DASH with significant signal to noise (SNR$>5$) in F160W. The SNR threshold ensures accurate GALFIT measurements for most of the galaxies. This sample excludes all galaxies from CANDELS in the 3D-HST morphological catalogs of \cite{vanderWel14} to avoid duplicate effort. This catalog can therefore be used directly alongside the COSMOS 3D-HST morphological catalog from \cite{vanderWel14}.  Masses in this sample are from the \cite{Muzzin13b} UVISTA catalog, and are derived using the FAST code \citep{Kriek09}. The mass-completeness limits for the UVISTA catalogs are shown in Table \ref{tab:complete} \citep[adapted from][]{Muzzin13b,Mowla19b}.

\begin{table}[h!]
    \centering
    \begin{tabular}{cc}\hline\hline
        $z$ & $\log(M/M_\odot)$ \\\hline
        $<0.5$ & 9\\
        0.5 - 1 & 9.6\\
        1 - 1.5 & 10.2\\
        1.5 - 2 & 10.6\\
        2 - 2.5 & 10.9\\
        2.5 - 3 & 11.1\\\hline
    \end{tabular}
    \caption{Table of 100\% mass-completeness limits for the UltraVISTA catalogs \citep{Muzzin13b}, adapted from Figure 3 of \cite{Mowla19b}.}
    \label{tab:complete}
\end{table}

Due to the limited spatial resolution of ground-based UVISTA observations (K-band FWHM$\sim0.8$\arcsec{} vs. HST F160W FWHM$\sim0.151$\arcsec{}), some sources cataloged as a single object in UVISTA will be deblended into two or more objects in COSMOS-DASH. Of the 51,586 galaxies in DASH, 2,345 are split into pairs or triplets when searching within the UVISTA FWHM. Adapting from \cite{Mowla19b}, for each deblended UVISTA ID we compute F160W total fluxes using GALFIT \citep{Peng02,Peng10}, making sure to mask the other deblended objects. The brightest galaxy is retained, with the stellar mass weighted by its flux relative to the total flux of all of the blended components:
\begin{align}
    M_{\star,i}~=~M_{\star,\mathrm{tot}}~\times~\frac{F_i}{F_\mathrm{tot}},
\end{align}
where $M_{\star,\mathrm{tot}}$ is the total mass of the blended object from the UVISTA catalog, $F_i$ is the model flux of the brightest component and $F_\mathrm{tot}$ is the total summed fluxes of all the components. The mass corrections are provided in the catalog of this sample (see Sec. \ref{sec:cat}).

The deblending status of the galaxies in the sample is indicated with a blending flag. HST-selected galaxies that have no close pairs are given a flag of 0, and are labeled ``not blended''. Galaxies whose components can be properly deblended and fit with GALFIT are flagged with 1, indicating they are ``deblended''. Deblended galaxies whose $U-V$ and $V-J$ colors do not change likely have robust masses and morphologies, but significant color changes may suggest that stellar masses and/or photometric redshifts are incorrect. Galaxies that remain ``blended'', meaning the deblending procedure failed, are given a flag of 2. Lastly, a flag of 4 is given to galaxies that are in the UVISTA catalogs, but have ``no coverage'' with COSMOS-DASH. This flag value is chosen to match with the ``no coverage'' GALFIT flag (see sec. \ref{sec:galfit}). Since later analyses in this work use rest-frame colors from UVISTA to classify galaxies as star-forming or quiescent, deblended galaxies with significantly differing colors must be excluded. The difference in the our analyses is negligible when all deblended galaxies (\texttt{flag\_deb}=1) are discarded. As such, we require a deblending flag of 0 in future analyses to avoid contamination from deblended galaxies with potentially incorrect colors, stellar masses, and redshifts.

\subsection{COSMOS-DASH Calibration Field}
We define a region of the mosaic as the ``calibration field''; this region contains roughly 12 overlapping pointings of CANDELS and COSMOS-DASH imaging. Since this calibration field contains both CANDELS F160W pointings and the new DASH pointings by design, we separate this data to create two versions:  exposures from (1) the COSMOS-DASH program only, and (2) the CANDELS program only. The CANDELS-only reduction overlaid on the DASH-only reduction is shown in the zoom panel of Fig. \ref{fig:mosaic}, with the regions where the two images overlap highlighted. The calibration field is useful for testing morphological fitting methods and comparing those results to the 3D-HST morphological catalogs, as it provides a reasonable sample size to run tests on and contains sources that have both DASH and CANDELS imaging. Through crossmatching the COSMOS-DASH HST-selected photometric catalog (see Sec. \ref{sec:sample}) with the 3D-HST morphological catalog \citep{vanderWel14}, we find 483 galaxies with significant F160W detection (SNR$>5$), comprising our calibration sample. Morphologies for this sample of calibration field galaxies are compared to the 3D-HST morphologies from \cite{vanderWel14}. This smaller sample allows for an in-depth error analysis and direct comparison of our measured errors relative to those of the corresponding galaxies in 3D-HST. The morphology and error analysis, as well as comparisons between the calibration field and 3D-HST, are discussed in more detail in the next section. For clarity, all figures using solely calibration field data are labeled in the top left.

\begin{figure}[hb!]
    \centering
    \includegraphics[width=\linewidth]{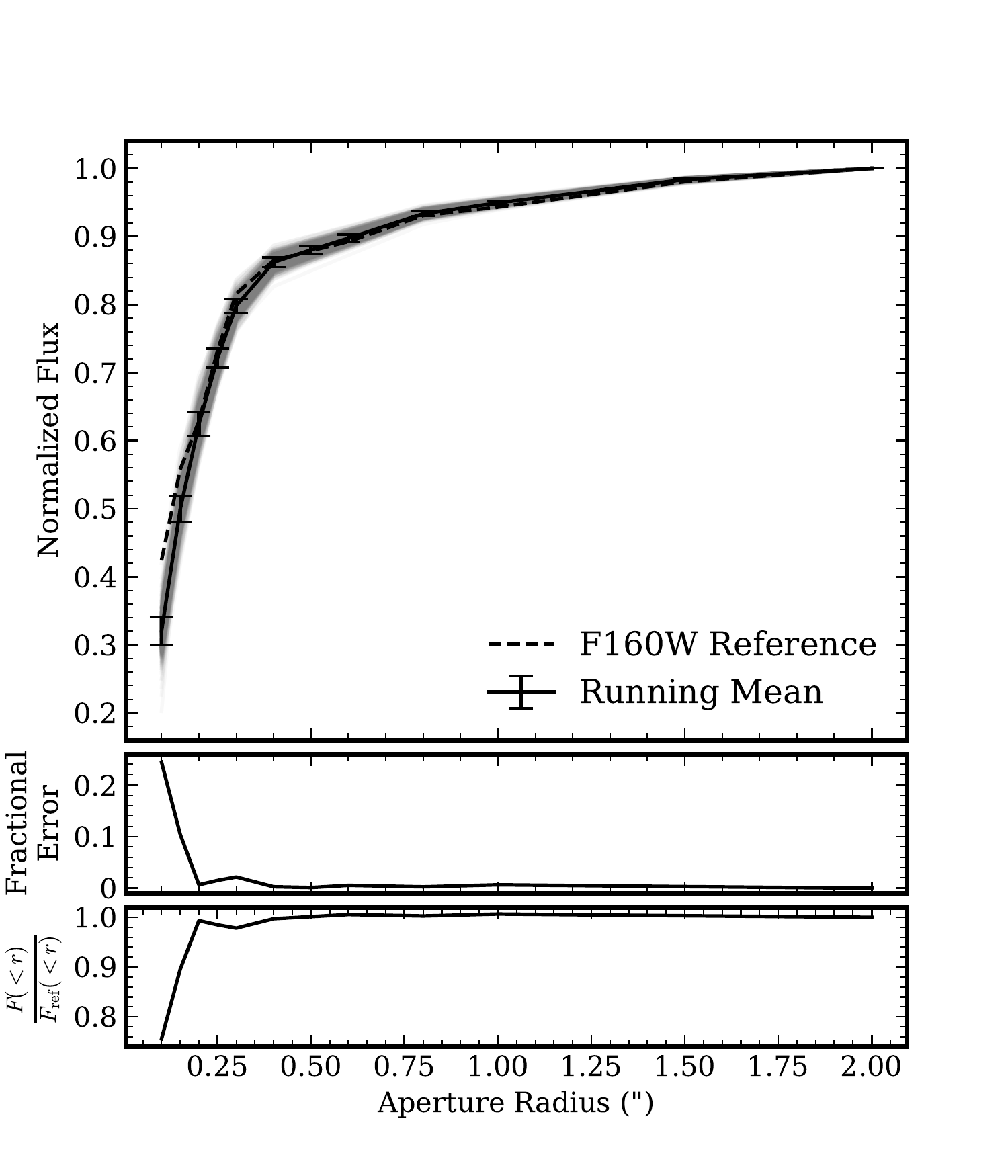}
    \caption{Normalized curve of growth for PSFs determined at positions of calibration field galaxies. The gray transparent lines are the growth curves for each of the 483 galaxies in the calibration field. The solid black line shows the running mean of the individual growth curves, with error from the standard deviation of the individual growth curves. The dashed line shows the F160W reference growth curve. The reference curve and each individual growth curve is normalized to unity at 2\arcsec{}. The middle panel shows the fractional error in the growth curve and the bottom panel shows the ratio between the reference curve and the running mean.}
    \label{fig:cog}
\end{figure}

\section{Analysis}\label{sec:methods}
\subsection{GALFIT Fitting Methodology}\label{sec:galfit}
We closely follow the fitting methods laid out in \cite{vanderWel12} and \cite{Mowla19b}, with a few notable changes made in our morphology fitting pipeline. We fit the list of COSMOS-DASH objects laid out in Section \ref{sec:sample}. A simple square cutout of each object and the corresponding location in the weight map is made using Montage v6.0\footnote{\url{http://montage.ipac.caltech.edu}}. The cutouts have a size equal to 7 times the semi-major axis Kron radius, which is measured when making the photometric catalog (Sec. \ref{sec:sample}). SExtractor is used to identify all sources in each cutout, masking all objects except for the galaxy of interest. This segmentation map is made in two ways: (1) using the SExtractor detection of the F160W cutout itself, and (2) from the ``global'' segmentation map created when building the deep, HST-selected catalog. The selection of which type of segmentation map we use is an important consideration which we will discuss more in Section \ref{sec:seg}. We also implement a position-dependent point spread function (PSF), the specifics of which will be discussed in Section \ref{sec:psf}. Lastly, we include a noise map with the same dimensions as the image cutout, determined identically to \cite{Mowla19b}. The noise calculations for DASH assume sky background is the dominant component and include Poisson noise.  Noise is calculated via
\begin{align}
    \sigma=\sqrt{\frac{|1+f|}{w}},
\end{align}
where $f$ is the image cutout and $w$ is the cutout weight image.

The image cutout, noise map, segmentation map, and position-dependent PSF are given to GALFIT \citep{Peng02,Peng10} which fits a S\'{e}rsic model. The free-parameters GALFIT changes are the magnitude ($m$), the semi-major axis effective radius ($r_\text{eff}$), the S\'{e}rsic index ($n$), axis ratio ($q$), the position angle (PA), and the central position ($x_0,y_0$). GALFIT can also fit a constant sky background to the input image; a discussion of background fitting is saved for Appendix \ref{sec:bkg}. Following \cite{vanderWel12} and \cite{Mowla19b}, initial guesses for the parameters are taken from SExtractor detection of the target galaxy. Boundary constraints are also placed on certain parameters: S\'{e}rsic index is held between 0.1 and 6, effective radius between 0.03\arcsec{} and 40\arcsec{} (0.3 and 400 pixels at the pixel scale of COSMOS-DASH), axis ratio between 0.0001 and 1, and magnitude between $-3$ and $+3$ magnitudes of the SExtractor magnitude, as in \cite{vanderWel12}. 

Galaxies are flagged for suspicious parameter values or for failed fits, similar to \cite{vanderWel12}. Objects that are fit without problems are assigned a flag of zero and are deemed ``good'' fits. ``Suspicious'' fits, sources whose GALFIT magnitudes deviate by more than $3\sigma$ (as determined in Section \ref{sec:err}) from the expected magnitude, are given a flag of 1. The expected magnitudes are measured in Section \ref{sec:phot} and are corrected for previously noted systematic offsets between GALFIT and SExtractor magnitudes, which range from roughly 0.1 dex at bright magnitudes to 0.3 dex at faint magnitudes \citep[see e.g.,][]{Haussler07,vanderWel12}. Galaxies that have a best-fit parameter at the constraint limits are considered ``bad'' and flagged with a value of 2. When GALFIT fails to converge at all, the fit is marked with a flag of 3 and we indicate the fit as ``failed''. Moreover, galaxies that have a negative aperture flux are also given a flag of 3, as the signal-to-noise ratio (SNR) is crucial in estimating parameter errors (see Sec. \ref{sec:err}). These galaxies account for only 0.5\% of all 51,586 galaxies with coverage. Lastly, we assign a flag of 4, differing from \cite{vanderWel12}, to galaxies in UVISTA that are not covered by COSMOS-DASH imaging. Though analysis is not possible for galaxies without COSMOS-DASH coverage, we still include these objects in order to present a complete list of IDs in the catalog (see Sec. \ref{sec:cat}). In total, 39.5\% of galaxies with coverage are ``suspicious'', 12.7\% are ``bad'', and 0.5\% are ``failed''. ``Suspicious'' galaxies are not necessarily poorly fit or reporting inaccurate magnitudes. In most cases, these are galaxies with very small GALFIT magnitude errors, so there is less of a deviation required for a flag to be applied. On average, ``good'' galaxies had a magnitude error that was 70\% larger than ``suspicious'' galaxies.

\begin{figure}[t!]
    \centering
    \includegraphics[width=\linewidth]{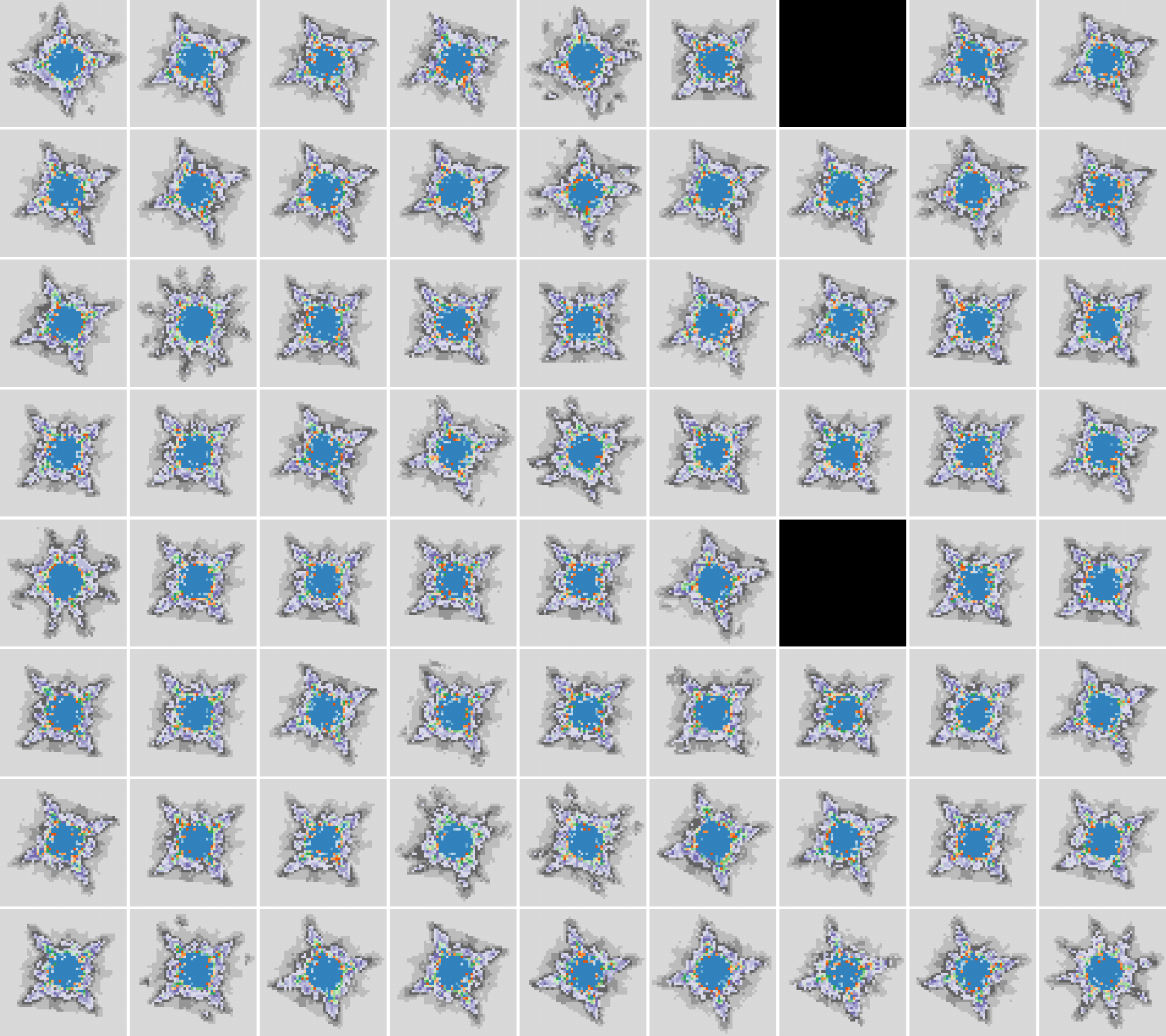}
    \caption{Grid of all 72 empirical PSFs for the COSMOS-DASH mosaic. Empirical PSFs are inserted in the frame of each exposure that covers a given position within the mosaic, and the PSFs are then drizzled with the same parameters as the science exposures. Black squares show regions where there is no PSF available because there are no dithering parameters.}
    \label{fig:psfs}
\end{figure}

\begin{figure}[hb!]
    \centering
    \includegraphics[width=\linewidth]{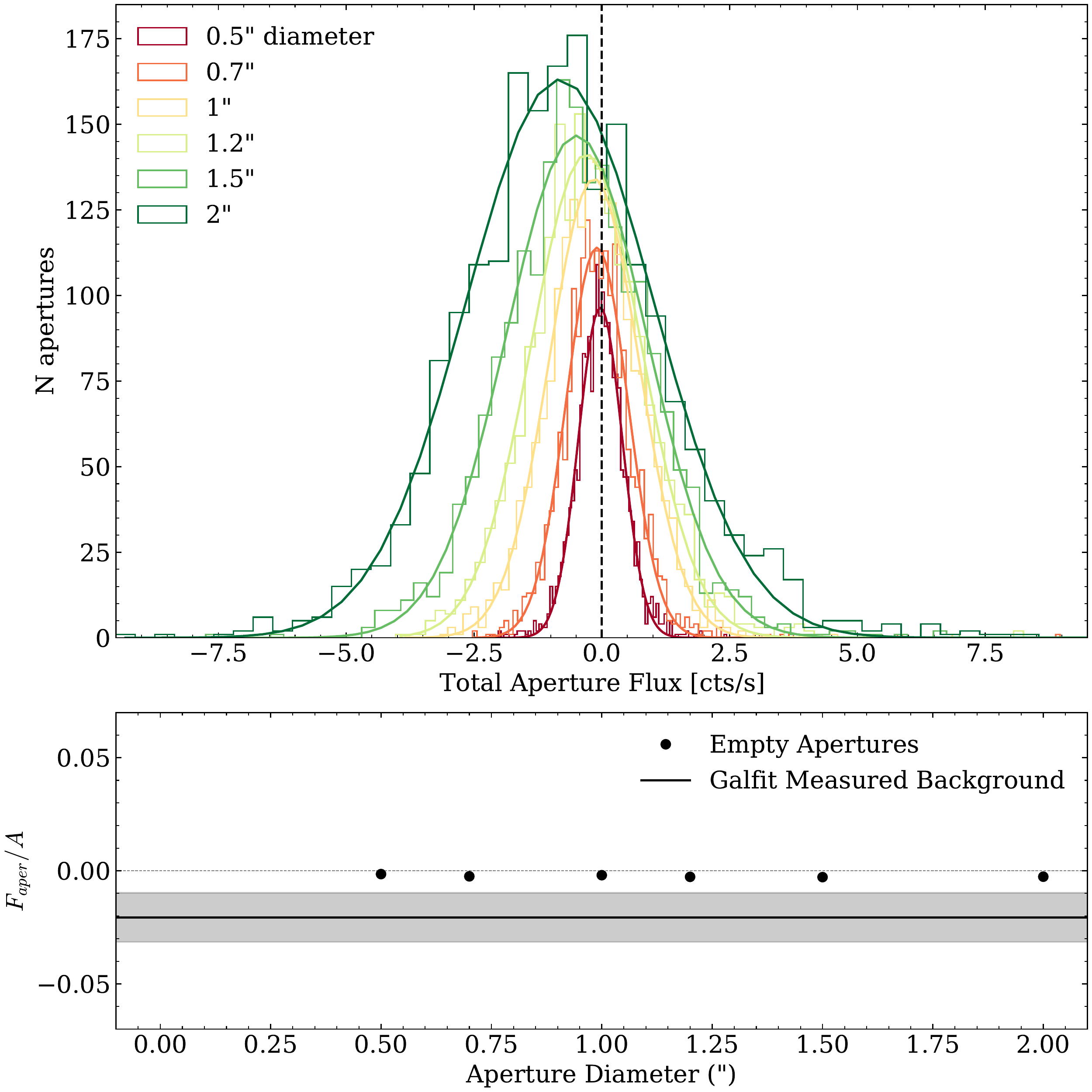}
    \caption{Empty aperture photometry on the background  COSMOS-DASH calibration field. The top panel shows histograms of the summed flux in different aperture sizes from empty regions in the image. The solid lines are the best fit Gaussians to these histograms. The bottom panel compares the means of these Gaussians divided by the number of pixels in the aperture (black points) to the mean background measured by GALFIT (solid line). The shaded region shown is the standard deviation in the GALFIT measured background.}
    \label{fig:empty_aper}
\end{figure}

\begin{figure*}[ht!]
    \centering
    \subfloat{\includegraphics[width=0.5\linewidth]{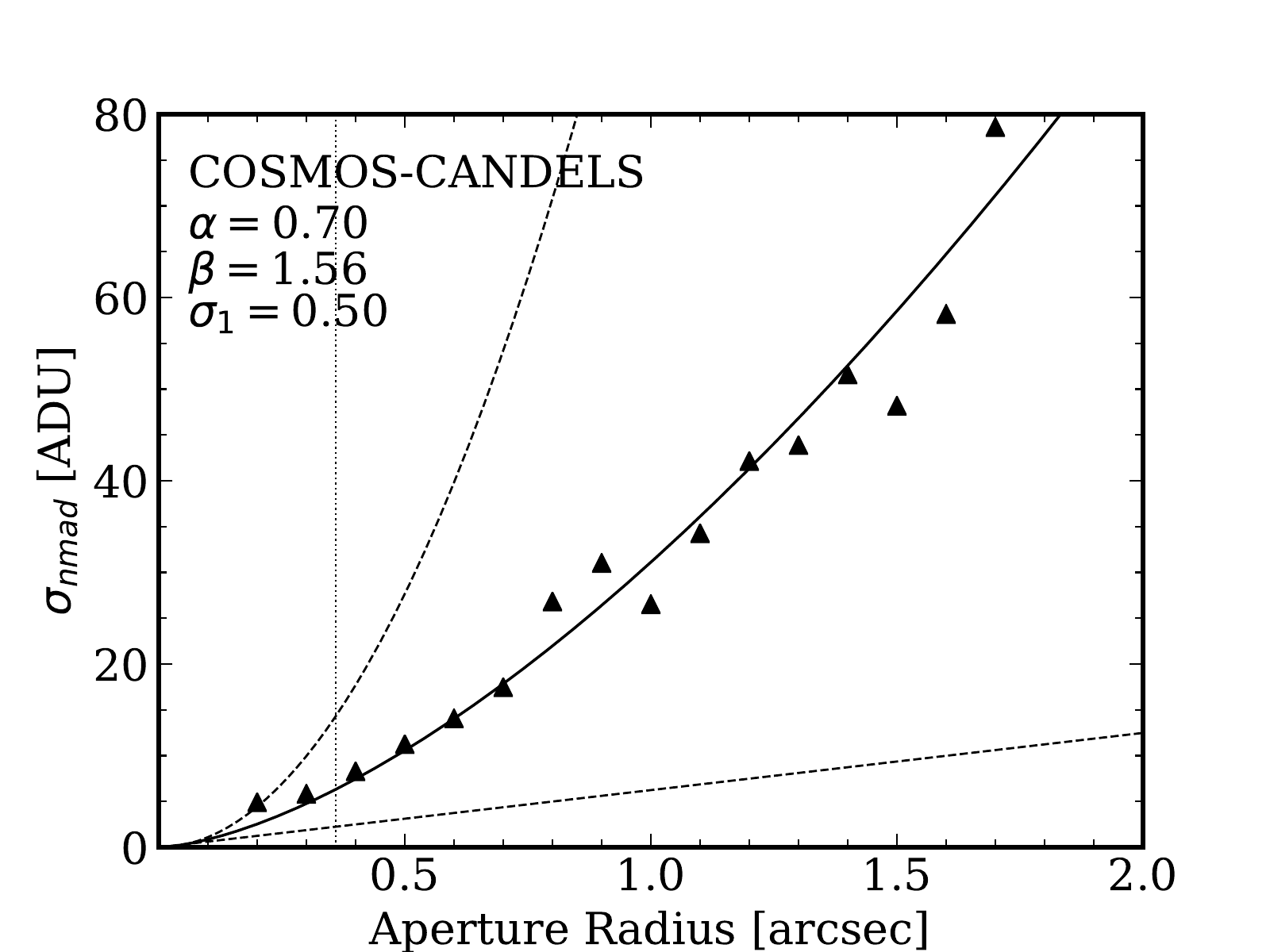}}
    \subfloat{\includegraphics[width=0.5\linewidth]{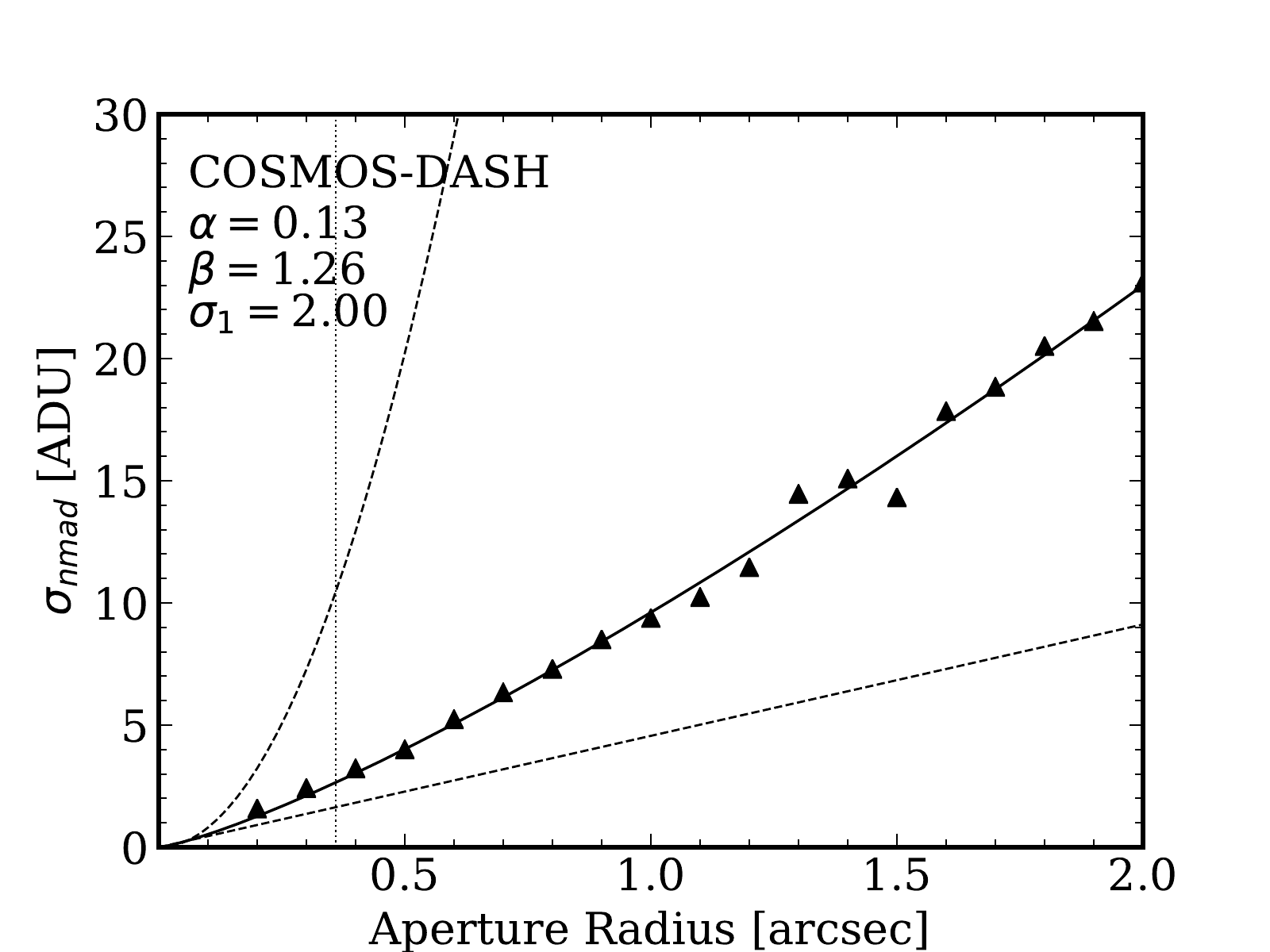}}
    \caption{Empty aperture errors for CANDELS-only (left) and DASH-only (right) reductions of the COSMOS-DASH mosaic. The measured best fit Gaussian width to the flux distribution of $2000$ apertures are indicated with triangles for a range of aperture sizes. The solid line indicates the best fit power law given by Eq. \ref{eqn:empty_ap}. Dashed lines show uncorrelated and perfectly correlated noise below and above the best fit line, respectively. The dotted line shows the PSF FWHM. The best fit power law parameters for each reduction are shown in the top left of each figure.} 
    \label{fig:aper_err}
\end{figure*}

\subsection{PSF Models}\label{sec:psf}
We obtain PSFs for COSMOS-DASH using Grizli\footnote{\url{https://grizli.readthedocs.io/}} to shift and drizzle HST empirical PSFs \citep{Anderson16} to (1) the location of every source, or (2) a coarser grid of positions throughout the mosaic when running GALFIT. Using the Grizli determined PSFs for all 483 galaxies in the COSMOS-DASH calibration field, we obtain growth curves for each galaxy (Fig. \ref{fig:cog}). The running mean and the standard deviation of the 483 individual growth curves is also calculated at each aperture size. We compare the running mean to the HST F160W reference curve and determine that there is no significant difference for aperture radii $\gtrsim0.25$\arcsec{}. Similarly, scatter in the position dependent PSFs is also negligible for a given aperture size. Moreover, there is no significant difference in structural measurements between galaxies fit with position dependent PSFs and those fit with PSFs determined over a larger area of the mosaic, even at $z>2$. This suggests that further corrections to the PSF at each individual location in the mosaic are unnecessary.

Due to the time it takes to determine each individual PSF and the minimal effect of individual position dependent PSFs, we alternatively opt for a set of empirical PSFs evaluated on a coarse grid across the mosaic. The 72 PSFs for the COSMOS-DASH mosaic are shown in Figure \ref{fig:psfs}. These are empirical PSFs that are altered to match the mosaic pixel scale and the dither grid at each position. In regions where there are no dithering parameters (due to the image being empty), the PSFs are also empty. These PSFs have an average FWHM of 0.2\arcsec{}, compared to 0.18\arcsec{} in CANDELS. For each galaxy, the closest PSF by RA and Dec is chosen, excluding empty PSFs. Further corrections (such as 2D interpolations) are not done due to the added computational time and negligible change in fit quality. This PSF is then used by GALFIT when fitting a 2D S\'ersic model like in \cite{vanderWel12} and \cite{Mowla19b}.

\subsection{Photometry and Noise Properties}\label{sec:phot}
We compute the flux density of all 51,586 galaxies within a 0.7\arcsec{} diameter aperture using SExtractor. This aperture size is chosen to maximize the SNR of HST observations \citep{Skelton14}. The Kron radius and Kron radius flux (AUTO flux) are also computed by SExtractor. The aperture flux is scaled to a total flux by correcting first to the flux within the Kron radius. We then interpolate the value of the growth curve (normalized to unity at 2\arcsec{}) at the Kron radius, and multiply the inverse of this value to obtain the total flux. If the AUTO flux is less than the aperture flux or the Kron radius is greater than 2\arcsec{}, then that scaling is kept at one following \cite{Skelton14}.

Photometric errors are estimated using an empty apertures approach \citep[e.g.,][]{Whitaker11,Skelton14}. For a diameter in the range of 0.2\arcsec{} to 2.1\arcsec{}, 2000 circular apertures are placed in empty regions of the noise-equalized, background-subtracted image. Apertures that overlap with objects based on the segmentation map or regions outside of the image are moved to different random coordinates until satisfactory. The distribution of the summed fluxes obtained for 6 aperture diameters ranging from 0.5\arcsec{} to 2\arcsec{} is shown in Figure \ref{fig:empty_aper} (top). The shifts in the Gaussian means indicate a small, per-pixel background, as shown in Figure \ref{fig:empty_aper} (bottom), which gets larger in magnitude as more pixels are included in the aperture. The figure also demonstrates that the widths of the best fit Gaussians increase with larger aperture sizes, which implies an increase in standard deviation with linear aperture size $N=\sqrt{A}$, where $A$ is the area of the aperture. This relation is modeled by a power law of the form
\begin{align}\label{eqn:empty_ap}
\sigma_\mathrm{nmad}=\sigma_1\alpha N^\beta,
\end{align}
where $\sigma_1$ is the standard deviation of the pixels of background pixels, $\alpha$ is the normalization, and $\beta$ is the power law index which falls between 1 and 2 \citep[see][]{Whitaker11}. A power law index of 1 indicates uncorrelated noise and a power law index of 2 indicates perfectly correlated noise. The combined COSMOS-DASH mosaic (outside the nominal CANDELS footprint)  contains shallower DASH-mode pointings alongside deeper standard HST observations (from archival imaging); we fit a separate power law for DASH- and standard-depth observations. The distinction in depth is determined from the weight maps, with a boundary set at a weight of 100, corresponding to a 5$\sigma$ depth of 26.7 ABmag (as calculated directly from the weight map). The resulting fit is shown in Figure \ref{fig:aper_err}. The standard-depth (CANDELS) imaging shows a higher level of correlation due to the finer spatial grid (when compared to DASH). To convert this standard deviation into a physical error, the noise equalization factor ($\sqrt{w}$, where $w$ is the weight) must be divided out.

To find the photometric error of each source, empty aperture error at the Kron radius is computed and added in quadrature to Poisson noise as follows:
\begin{align}
    \sigma_\text{Kron}^2=\left(\frac{\sigma_1\alpha\left(\pi R_\text{Kron}^2\right)^\frac{\beta}{2}}{\sqrt{w}}\right)^2+\frac{F_\text{AUTO}}{g},
\end{align}
where $w$ is the weight of the source, $F_\text{AUTO}$ is the Kron radius flux of the source, and $g$ is the effective gain (detector gain times exposure time) of the data. This is then scaled to a total error using the growth curve in the same way as the flux. 

\begin{figure*}[ht!]
    \centering
    \includegraphics[width=\linewidth]{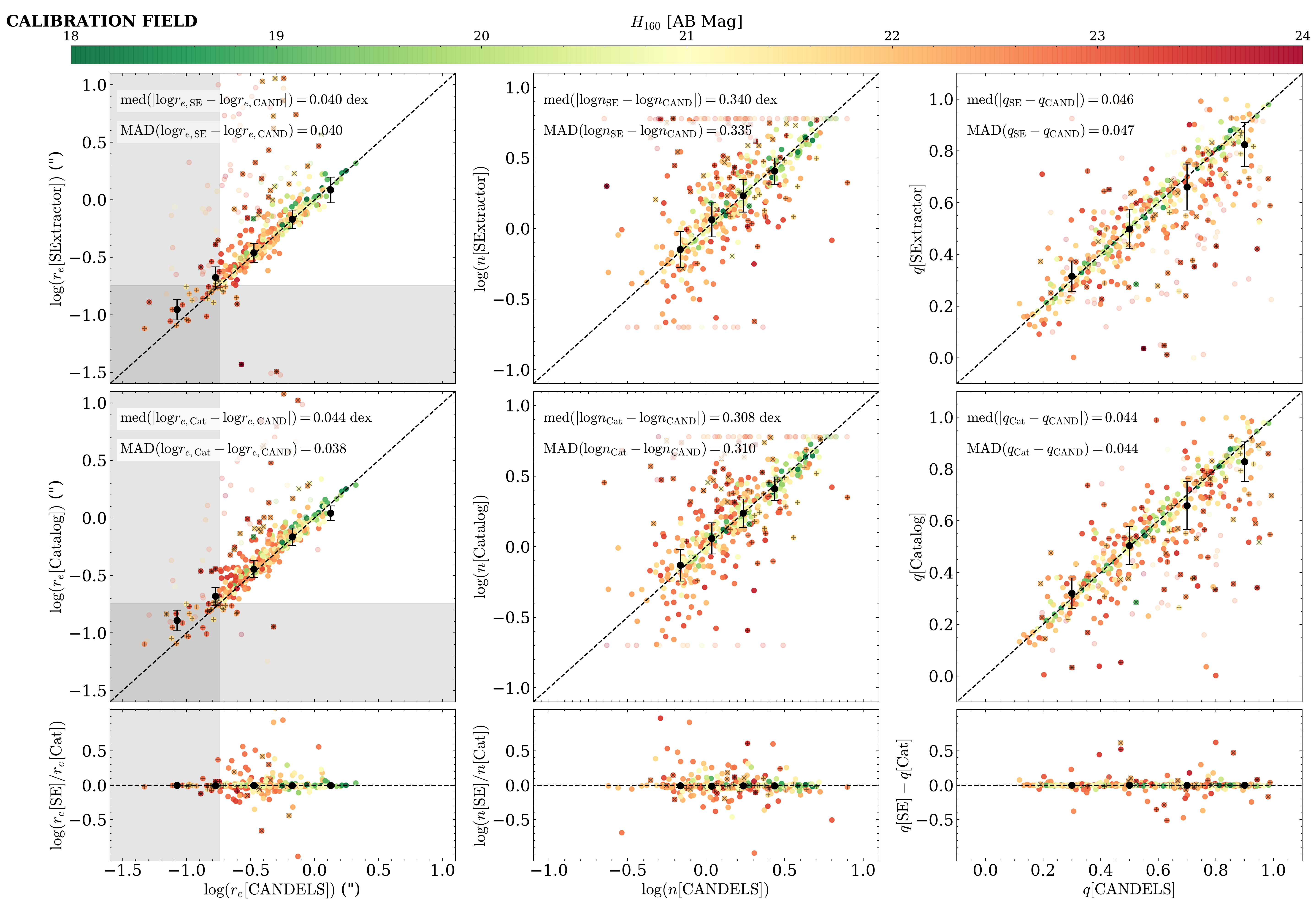}
    \caption{Scatter in the best fit S\'ersic parameters obtained using a deep-detection (Catalog) segmentation map is slightly smaller than with a DASH-depth (SExtractor) segmentation map. The top and middle rows show the effective radii, S\'ersic indices, and axis ratios of the best fit S\'ersic models determined with the DASH-depth segmentation map (top) and with the catalog segmentation map (middle), respectively, compared to those from CANDELS/3DHST \citep{vanderWel12}. Transparent points are those with \texttt{flag}=2 parameter values (see Section \ref{sec:galfit}). The bottom row shows the difference in the parameters when measured both with and without fitting background. The large black points are the running median in each bin, with errorbars showing the scatter determined by the median absolute deviation (MAD) of the bin. The color scale indicates the $H_{160}$ magnitude of the source in AB magnitudes. Points with black x's are those that are greater than 0.3 dex from the one-to-one line in radius. The grey shaded region shows radii less than the PSF FWHM and points with black +'s are sources whose best fit radius falls in this region. The median and MAD of the offset between parameters derived with the two segmentation maps and CANDELS is shown in the each panel of the top and middle row.}
    \label{fig:seg_comp}
\end{figure*}

\begin{figure*}[ht!]
    \centering
    \includegraphics[width=\linewidth]{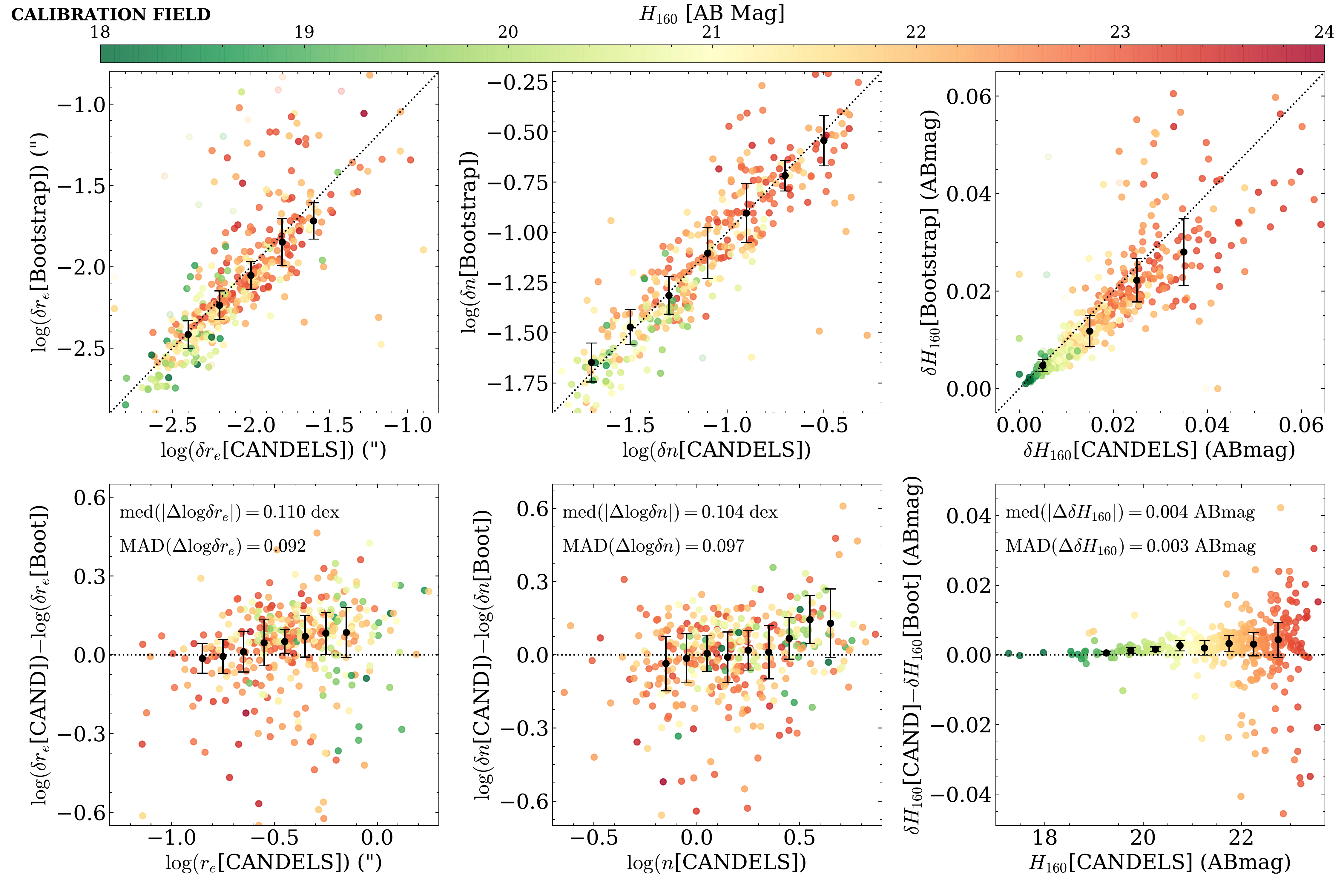}
    \caption{GALFIT parameter errors of calibration field galaxies derived from bootstrapping are in agreement with errors the CANDELS/3D-HST catalog. The top row shows, from left to right, the bootstrapping error in effective radius, S\'ersic index, and magnitude plotted against the same errors from the CANDELS/3D-HST. Bootstrapping is run on galaxies in the CANDELS-only reduction of the calibration field. The dashed line shows the one-to-one line. The bottom row shows the difference between the CANDELS/3D-HST and bootstrapping errors plotted against the CANDELS/3D-HST effective radius, S\'ersic index, and magnitude. The running median in each panel is shown with black points, and error bars are determined with the median absolute deviation. The dashed line shows a difference of zero. The color scale indicates the AB magnitude of each galaxy, as reported in the COSMOS-DASH catalog. The median and MAD of the offset between bootstrapping and CANDELS errors are indicated in each of the bottom panels.}
    \label{fig:bootstrapping}
\end{figure*}

\subsection{Segmentation Map}\label{sec:seg}
Due to the uncertain depth and structural fidelity of DASH imaging, we choose to mask objects in lieu of simultaneously fitting neighboring objects with GALFIT. This is done to circumvent any issues with poorly modeled neighboring objects impacting the structural fits to the target galaxy. When fitting structural parameters with masking, the effect of different size segmentation maps resulting from detection images of varying depth becomes an important consideration. 
The deeper images used in the COSMOS-DASH photometric catalog are able to detect more of the fainter, extended flux of galaxies. Thus, the segmentation map from these images is larger than when detected in DASH-depth imaging. When objects surrounding the target galaxy are masked by GALFIT, a larger area mask of surrounding objects could remove flux from the faint, extended wings of large S\'ersic index target galaxies, especially in crowded regions of the mosaic. This could suggest that sources fit using segmentation maps built from DASH-depth imaging have systematically larger S\'ersic indices than those with deeper detection images.

To test the effect of two different size segmentation maps on the best-fit parameters, we first use the COSMOS-DASH photometric catalog segmentation map, which is made from deep imaging: this segmentation map uses all available HST imaging that overlaps with the HST-WFC3 F160W observations of COSMOS-DASH field, most notably HST-ACS F814W data \citep{Scoville07,Koekemoer07} and HST-WFC3 F160W and F125W data from CANDELS \citep{Koekemoer11}. A cutout of this catalog segmentation map is made with Montage and given to GALFIT, similar to the image cutout. Next we generate segmentation maps from DASH-depth data. To do this, SExtractor \citep{Bertin96} is run on the individual COSMOS-DASH cutouts (Sec. \ref{sec:galfit}), similar to \cite{Mowla19b}, and the resulting segmentation maps are used with GALFIT. We run the GALFIT pipeline on the 483 calibration field galaxies using both segmentation maps. 

In Figure \ref{fig:seg_comp}, we compare the measured morphologies of both scenarios. 
The first row compares parameters measured using the DASH-depth segmentation map to the morphologies of corresponding galaxies in the CANDELS/3D-HST morphological catalog. The second row makes a similar comparison showing parameters measured using the deep-detection (catalog) segmentation maps. 
Galaxies with a flag value $\geq3$ are excluded from the plot and objects with a flag of 2 are indicated with higher transparency and excluded from the bottom row of the figure. For the comparison, CANDELS/3D-HST galaxies with flags $\geq2$ \citep[see,][]{vanderWel12} are also excluded. Objects that deviate significantly in radius (more than 0.3 dex) from CANDELS/3D-HST are indicated with black x's. If a source has a measured radius less than the PSF FWHM (0.18\arcsec{}), it is marked with a black ``+'', as radius measurements may be unreliable below the FWHM.

On average, both scenarios have morphologies consistent with those from the CANDELS/3D-HST catalog. We observe a 22\% larger scatter in the measured radii and an 11\% larger scatter in S\'ersic indices for fits obtained with the DASH-depth segmentation map. Given the factor of two decrease in standard deviation when using the deeper-detection segmentation map, we adopt the deep-detection segmentation map for masking, as this best represents the full extent of detected sources. In the case where deeper data isn't available, it may be advisable to ``grow'' the DASH-only segmentation maps in size, though this was not implemented in this analysis due to the existence of deeper data throughout COSMOS from HST-ACS F814W \citep{Scoville07,Koekemoer07}.

\begin{figure*}[ht!]
    \centering
    \includegraphics[width=\linewidth]{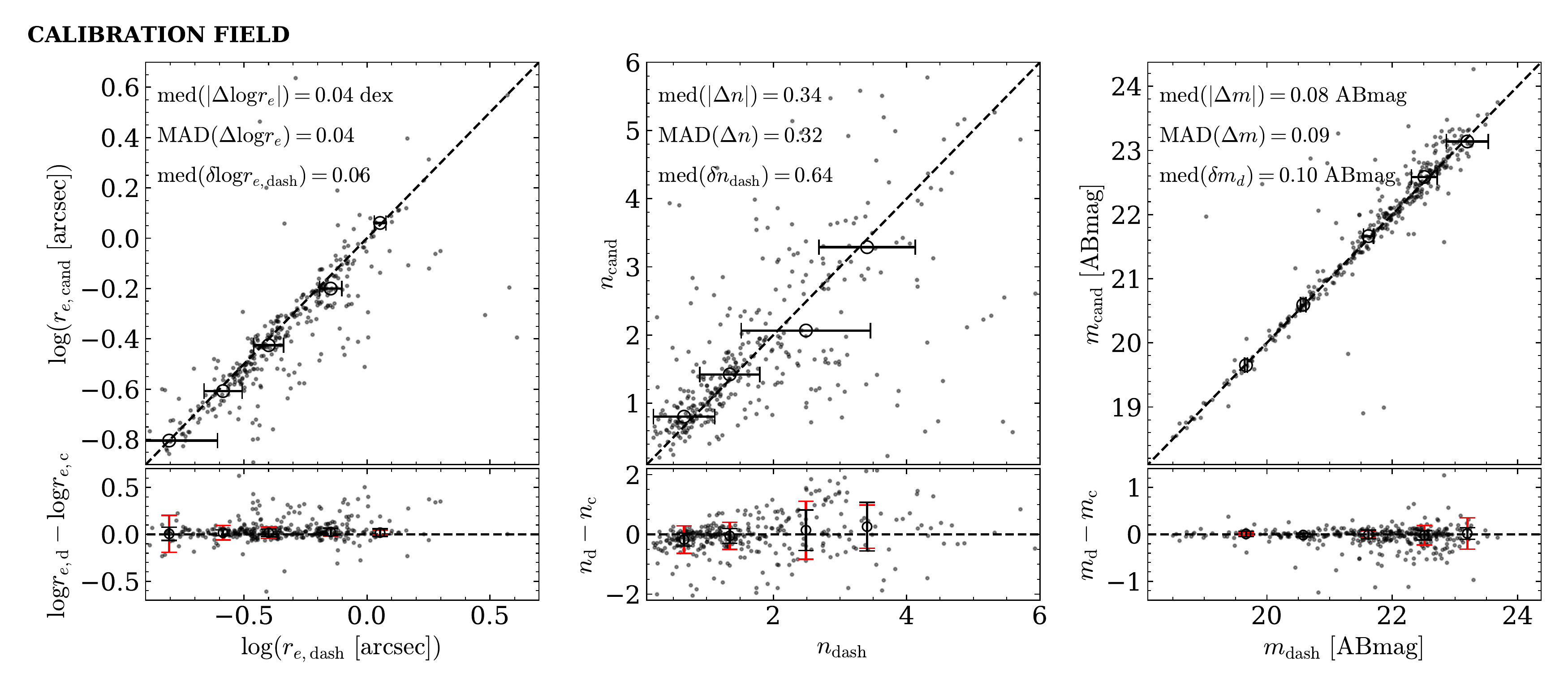}
    \caption{Size, S\'ersic index, and magnitude comparison between calibration field galaxies in CANDELS/3D-HST and COSMOS-DASH, showing overall agreement between parameter values and random uncertainties consistent with scatter. In the top row, CANDELS/3D-HST parameters with GALFIT flags $\leq1$ \citep[see ][]{vanderWel12} are plotted against COSMOS-DASH paramters with GALFIT flags $\leq1$. The running median and one-to-one line are shown with open circles and dashed lines, respectively. $1\sigma$ error bars show the average parameter error (from bootstrapping) in each running median bin. The bottom row shows the difference between CANDELS/3D-HST and COSMOS-DASH structural parameters plotted against the COSMOS-DASH parameters. The open circles again indicate the running median, while the dashed line shows a difference of zero. The black error bars are the median absolute deviation in the running median, while the red error bars show the average parameter error. The median and MAD of the difference in the structural parameters, as well as the average parameter error is also displayed in the top left of each panel.}
    \label{fig:comp}
\end{figure*}

\subsection{Bootstrapping and Error Estimation}\label{sec:err}
In order to discern where DASH-depth morphologies become unreliable, consistent parameter errors must be measured. To estimate the error in each GALFIT parameter, we run a bootstrapping analysis on the 483 galaxy calibration field sample. A random array the same dimensions as the image cutout is drawn from a normal distribution. The random perturbation is multiplied element-wise by the noise image (see Section \ref{sec:galfit}) and by a scalar empty aperture error $\sigma(A)$, see Eq. \ref{eqn:empty_ap}, where $A$ is the aperture area in square pixels. This error-scaled random perturbation is added to the image, which is then fit using GALFIT. The initial parameter guesses given to GALFIT are the best fit results of the GALFIT routine run without perturbation. This is repeated for 50 iterations with a different random perturbation for each iteration. The mean of the 50 best fit parameters are taken as the parameter value and the standard deviation is used as the error in the parameter measurement.

We test the bootstrapping method on the calibration field sample of galaxies using the CANDELS-only reduction. This allows us to make direct comparisons to the CANDELS/3D-HST morphological catalog errors, as these analyze the same sources at the same depth.  In lieu of extrapolating to a single pixel, we adopt the empty aperture error within 3 pixels when running the bootstrapping simulations. We find this is consistent with CANDELS/3D-HST errors within $1\sigma$. Figure \ref{fig:bootstrapping} shows this comparison of CANDELS-only calibration field errors (from COSMOS-DASH) to CANDELS/3D-HST errors. Brighter sources have smaller errors, as these are more easily fit. There is also strong agreement between the photometric catalog magnitudes and the morphological catalog magnitudes (Fig. \ref{fig:bootstrapping}, bottom right). Galaxies with smaller errors have less scatter around the one to one line than those with larger errors, which is expected as well. The parameter error difference is also less than 0.3 dex for most galaxies in both radius and S\'ersic index and not significant in magnitude error for all but the faintest galaxies. This is confirmed by the median offsets between the errors, as well as the scatter in this offset (bottom panels, Fig. \ref{fig:bootstrapping}). 

We also directly compare the parameter values of calibration field galaxies in CANDELS/3D-HST and COSMOS-DASH in Figure \ref{fig:comp}. The running median (open circles) is in agreement with a one-to-one relationship within the error bars, which are given by the median absolute deviation. Moreover, the random uncertainties from bootstrapping are consistent with the scatter in the CANDELS/3D-HST and COSMOS-DASH relation, as shown by the red and black $1\sigma$ error bars, respectively. This is also reflected by quantitative measurements of the difference between the parameters and parameter error that are indicated in the top left of the panels in Figure \ref{fig:comp}. This indicates that there are no systematic effects in both our structural measurements and our error analysis.

\begin{figure*}
    \centering
    \includegraphics[width=\linewidth]{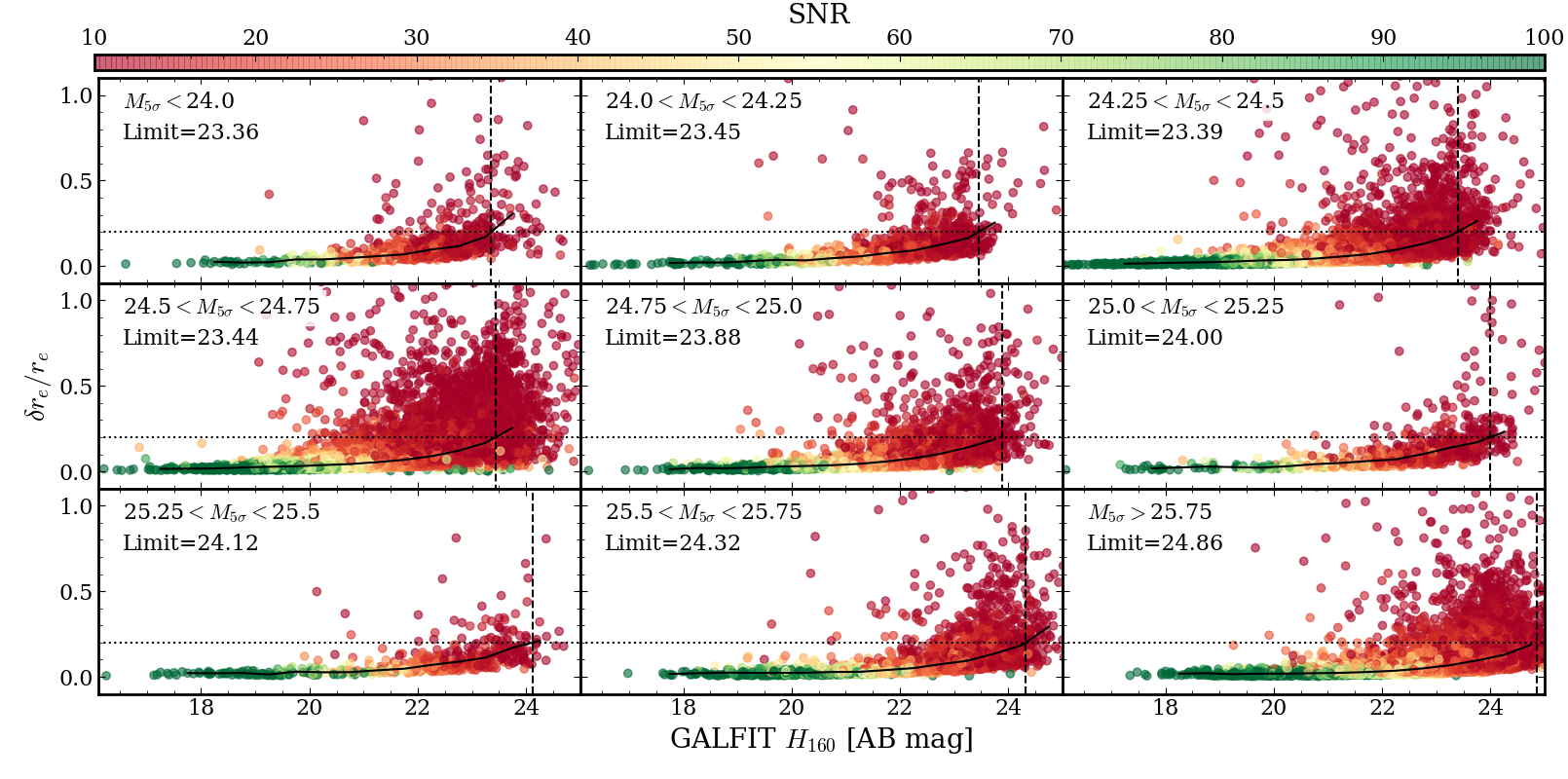}
    \includegraphics[width=\linewidth]{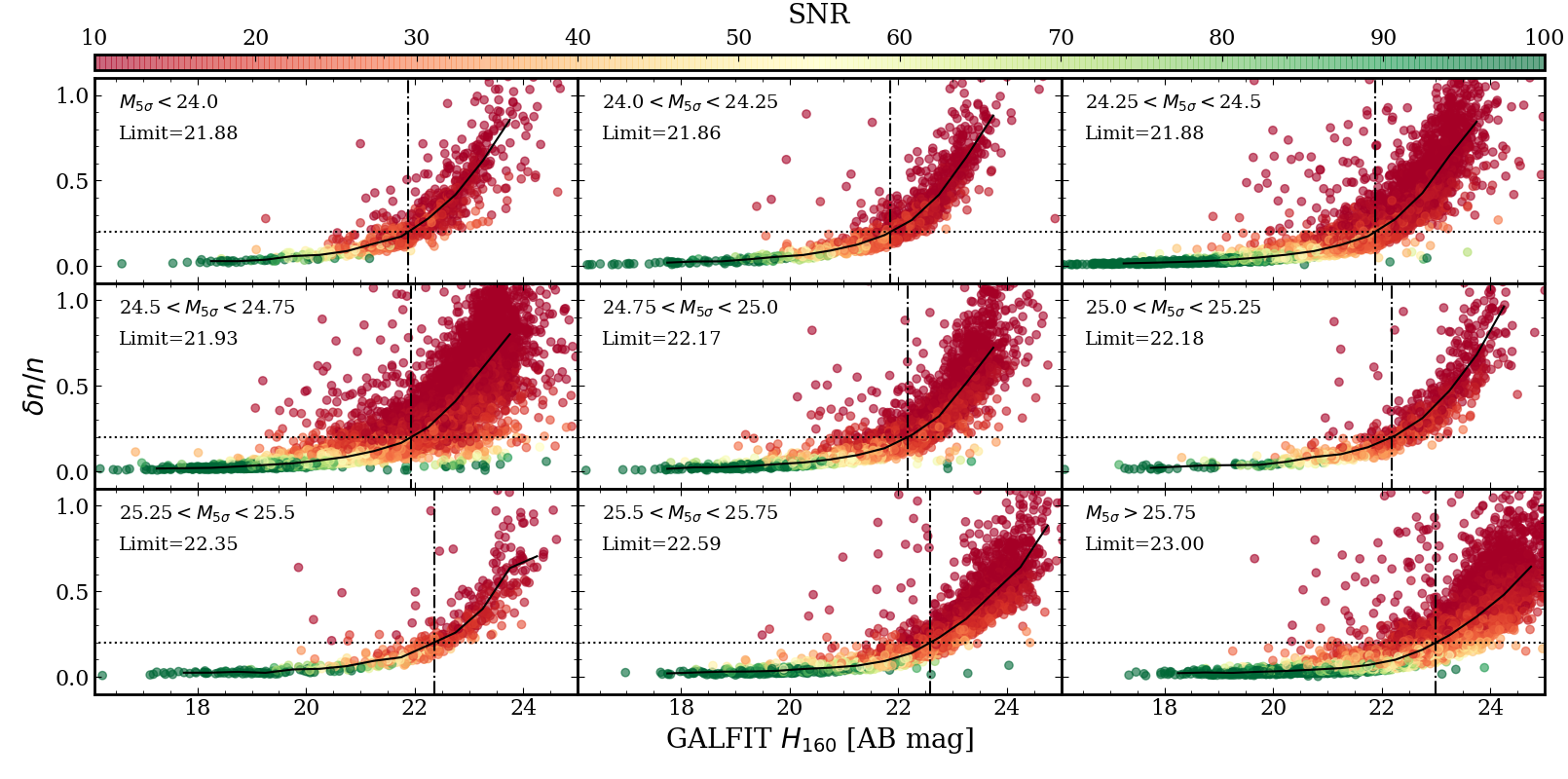}
    \caption{Evolution of relative uncertainty in effective radius (top) and S\'ersic index (bottom) with increasing depth, resulting in the parameter magnitude limits of Figure \ref{fig:param_lim}. For each depth bin ($m_{5\sigma}$) given in the top left of each panel, relative errors in effective radius and S\'ersic index (Sections \ref{sec:galfit} and \ref{sec:err}) are plotted against the best fit GALFIT magnitude. A $3\sigma$ clipped running mean is shown with a solid line. A relative error of 20\% is marked by a dotted line. The running median is linearly interpolated to find the magnitude where the fractional error reaches 20\%. This value is indicated with a dashed vertical line and explicitly stated by the limit in the top left of each panel. The color bar shows the SNR of each galaxy as computed in Section \ref{sec:phot}.}
    \label{fig:err}
\end{figure*}
\begin{figure}[t!]
    \centering
    \includegraphics[width=\linewidth]{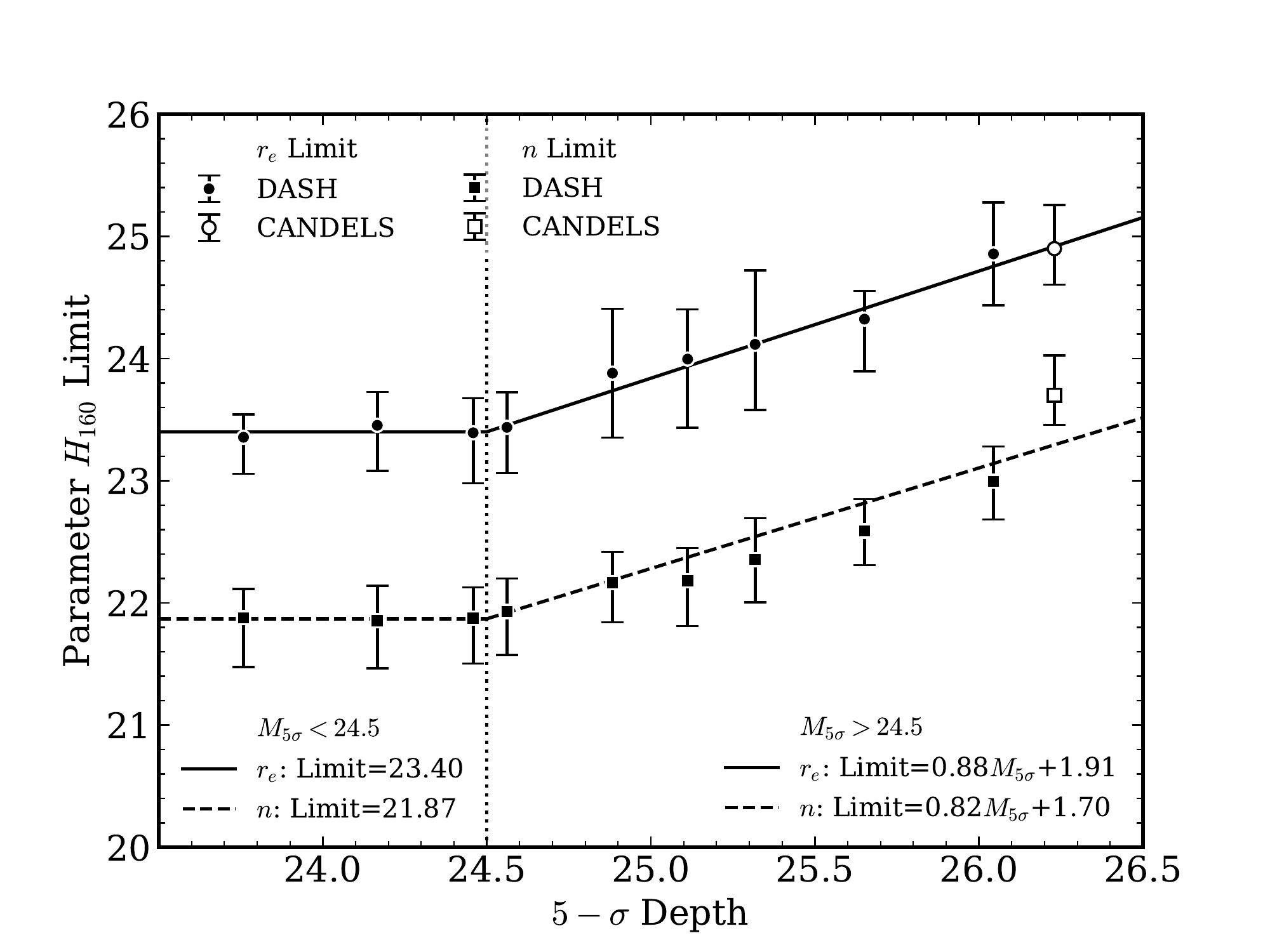}
    \caption{Magnitude limit-depth relation for effective radius and S\'ersic index used in generating parameter robustness flags in the COSMOS-DASH morphological catalog. The GALFIT $H_{160}$ limits derived in Section \ref{sec:lim} and Figure \ref{fig:err} are compared to the $5\sigma$ depths computed in Section \ref{sec:lim}. Effective radius limits are shown with circles and S\'ersic index limits are shown with squares. Limits from COSMOS-DASH (this study) are indicated by filled points while open points indicate values derived from the CANDELS/3D-HST morphological catalog \citep{vanderWel14}. The expression describing the relation between the parameter limits and the magnitude depths is a piecewise function that is constant for depths $<24.5$ and linear for depths $>24.5$. The dotted line indicates a $5\sigma$ depth of 24.5, illustrating the two regimes of the relation. The resulting best fit constant and linear relations are shown in the bottom left and right of the figure and the curves are indicated with a solid line for radius and a dashed line for S\'ersic index.}
    \label{fig:param_lim}
\end{figure}

Computational constraints prevent us from determining parameter errors using bootstrapping for all 51,586 galaxies in our sample. Instead, we adapt the empirical error estimation approach from Section 5.1 of \cite{vanderWel12} that will use our calibration field errors obtained from bootstrapping. For each galaxy in the COSMOS-DASH morphological catalog with GALFIT flag $\leq1$, we find the 25 most structurally ``similar'' galaxies in the calibration field. This is done by computing the ``position'' of each galaxy in the three-dimensional parameter space given by
\begin{align}
    \Vec{p}_i=(m_i/\sigma(\Vec{m}),~\log n_i/\sigma(\log\Vec{n}),~\log r_i/\sigma(\log\Vec{r})),
\end{align}
where $\sigma$ is the standard deviation in a parameter for the entire sample set. 

We compute $\Vec{p}_i$ for each galaxy in the calibration field and use the bootstrapping results to calculate the signal-to-noise ratio (SNR) equalized parameter error vector:
\begin{align}
    \Vec{\delta}_i=(\delta m_i,~\delta\log n_i,~\delta\log r_i,~\delta q_i,~\delta\text{PA}_i)\cdot(S/N)_i,
\end{align}
where $(S/N)_i$ is the SNR of the galaxy. The multiplication by SNR removes first-order SNR dependence. The set of $\Vec{p}_i$ and $\Vec{\delta}_i$ can then be used as a representative sample of galaxies and errors from which we can estimate the parameter error of similar galaxies. 

For each galaxy in the full mosaic with position $\Vec{p}_i$, we compute the ``distance'' to each galaxy in the calibration field via $\Vec{p}_i-\Vec{p}_j$, where $\Vec{p}_j$ is the position of galaxies in the calibration field. Following \cite{vanderWel12}, the 25\footnote{We find no significant change in uncertainties if the number of galaxies is reduced to 10 or increased to 50, in agreement with \cite{vanderWel12}.} most similar galaxies (i.e. galaxies with the shortest distance) in the calibration field are selected and $\Vec{\delta}_i$ is computed by taking the average of $\Vec{\delta}_j$. We then divide $\Vec{\delta}_i$ by $(S/N)_i$ to properly normalize the parameter error. Log errors are converted to linear errors in the standard way, i.e. $\delta r=\ln(10)~r~\delta\log r$, and these errors are given in the morphological catalog. The relative errors of the full DASH sample are shown in Figure \ref{fig:err} compared to the corresponding GALFIT magnitudes.

\subsection{Parameter Limits}\label{sec:lim}
Fainter and more noisy sources are expected to yield less robust structural diagnostics than bright, high SNR galaxies. Thus, we can directly relate the depth of the data to the magnitude limit at which morphologies become unreliable. We measure the depth of the entire COSMOS-DASH mosaic (Fig. \ref{fig:mosaic}) using empty apertures on the noise-equalized image (see Section \ref{sec:phot}). We separately place 2000 apertures in empty regions of DASH and CANDELS depth areas of the full mosaic. The width of the best-fit gaussians for both aperture flux distributions is computed and used with the weight map to find a $5\sigma$ depth ($m_{5\sigma}$). 

For CANDELS depth data ($m_{5\sigma}\sim26.5$), galaxies brighter than $H_{F160W}\sim24.5~(23.5)$ had average fractional errors of 20\% or lower for the parameters $m,~r_e,$ and $q$ ($n$) \citep[see also][]{vanderWel12}. We find similar values for the CANDELS/3D-HST morphological catalog (24.9 and 23.7, respectively) using a $3\sigma$ clipped running mean. A similar analysis is done on the 51,586 DASH galaxies, shown in Figure \ref{fig:err}. Fractional errors for both effective radius and S\'ersic index are computed using the errors derived in Section \ref{sec:err} and the best fit parameters from GALFIT (\ref{sec:galfit}). We compute the $5\sigma$ depth of each object and separate each galaxy in to 0.25 AB magnitude bins based on the depth of the image at their location. The fractional errors are plotted against the GALFIT best-fit magnitudes and color-scaled by the total SNR (i.e. the aperture SNR scaled to total using the curve of growth). A running mean over the respective GALFIT magnitudes is computed for each depth bin with outliers $>3\sigma$ clipped (solid line in Fig. \ref{fig:err}). We interpolate the running mean to find the magnitude at which it equals a fractional error of 0.2 (marked with a dashed-dotted line in Fig. \ref{fig:err}) and quote that magnitude in the top left of each panel. 

These values are then plotted against the corresponding $5\sigma$ depth in Figure \ref{fig:param_lim} (solid points) along with the limits from CANDELS/3D-HST (open points). We separate the limits into two regimes, shown with a dotted line in Fig. \ref{fig:param_lim}: $m_{5\sigma}<24.5$, where the parameter limit is constant with changing depth, and $m_{5\sigma}>24.5$, where there is a linear relationship between depth and magnitude limit. For the former, we measure the mean parameter limit for all bins with $m_{5\sigma}<24.5$ and use this as the constant limit for this regime\footnote{While all of these exposures are taken with the DASH method, varying zodiacal light impacts the noise level and pointing depth.}. For $m_{5\sigma}>24.5$, ensuring continuity at $m_{5\sigma}=24.5$, we fit a linear relation to the data. The resulting depth-parameter limits are
\begin{align}\label{eqn:rlim}
    m_{lim,r}&=
    \begin{cases}
        23.40&[m_{5\sigma}<24.5],\\
        0.88m_{5\sigma}+1.91&[m_{5\sigma}>24.5],
    \end{cases}
\end{align}
for effective radius and 
\begin{align}\label{eqn:nlim}
   m_{lim,n}&=
    \begin{cases}
        21.87&[m_{5\sigma}<24.5],\\
        0.82m_{5\sigma}+1.70&[m_{5\sigma}>24.5],
    \end{cases}
\end{align}
for S\'ersic index. They are also shown as solid and dashed lines, respectively, in Figure \ref{fig:param_lim}. These limits can be useful for future planned observations using DASH. If a survey with an expected imaging depth needs robust morphologies for galaxies of a certain magnitude, these magnitude limits can be used as checks to ensure this is feasible.  

\begin{deluxetable*}{lll}
\centering
\tabcolsep0.3in
\tablewidth{\linewidth}
\tablecaption{Summary of Morphological Catalog Contents\label{tab:header}}
\tablecolumns{3}
\tablehead{
\colhead{Column No.}&
\colhead{Column Title}&
\colhead{Description}
}
\startdata
1 & ID & Object identifier from UVISTA catalogs of \cite{Muzzin13a} \\
2,3 & RA,DEC & Right ascension and declination (J2000; decimal degrees) \\
4 & flag & GALFIT flags (see Sec. \ref{sec:galfit}); 0=good, 1=suspicious, 2=bad, 3=failed, 4=no coverage\\
5 & use & General use flag (see Sec. \ref{sec:cat}); 1=GALFIT flag<2, $r_e>$FWHM\\
6 & mag & GALFIT best-fit magnitude\\
7 & dmag & Uncertainty in GALFIT magnitude (see Sec. \ref{sec:err})\\
8 & re & GALFIT best-fit effective (half-light) radius in arcsec\\
9 & dre & Uncertainty in GALFIT effective radius in arcsec\\
10 & n & GALFIT best-fit S\'ersic index\\
11 & dn & Uncertainty in GALFIT S\'ersic index\\
12 & q & GALFIT best-fit axis ratio\\
13 & dq & Uncertainty in GALFIT axis ratio\\
14 & pa & GALFIT best-fit position angle\\
15 & dpa & Uncertainty in GALFIT position angle\\
16 & kron & Kron radius from SExtractor in arcsec\\
17 & f\_F160W\_auto & F160W AUTO flux (SExtractor measured flux within Kron radius); zeropoint=25\\
18 & e\_F160W\_auto & Error in F160W AUTO flux; zeropoint=25\\
19 & f\_F160W\_tot & F160W total flux (scaled from AUTO flux); zeropoint=25\\
20 & e\_F160W\_tot & Error in F160W total flux (scaled from AUTO flux error); zeropoint=25\\
21 & snr & Total signal-to-noise ratio (f\_F160W\_tot/e\_F160W\_tot)\\
22,23 & flag\_limit\_r, flag\_limit\_n & Parameter robustness flags (see Sec. \ref{sec:cat}); 0=not robust, 1=robust\\
24 & flag\_deb & Deblending flag (see Sec. \ref{sec:sample}); 0=not blended, 1=deblended, 2=blended, 4=no coverage\\
25 & Mcorr & Mass correction for deblended galaxies (see Sec.\ref{sec:sample})\\
26 & 5\_sigma\_depth & Total $5\sigma$ depth measured from COSMOS-DASH mosaic (see Sec. \ref{sec:lim})\\
27 & chi & GALFIT chi-squared\\
28 & chi\_nu & GALFIT chi-squared per degree of freedom
\enddata
\end{deluxetable*}

\section{Morphological Catalog}\label{sec:cat}

This paper is accompanied by the public release of the COSMOS-DASH morphological catalog, available at MAST as a High Level Science Product via \dataset[10.17909/T96Q5M]{\doi{10.17909/T96Q5M}}. Of the 262,615 objects identified in the UVISTA catalogs of \cite{Muzzin13a}, 51,586 have coverage with the COSMOS-DASH F160W observations. The catalog contains object IDs from \cite{Muzzin13a}. Best-fit GALFIT parameters and errors from this study are also given in the catalog, when available, as well as photometric fluxes and errors measured in Section \ref{sec:phot} and $5\sigma$ depths measured in Section \ref{sec:lim}. The catalog fluxes can be converted into total magnitudes using a photometric zeropoint of 25. The information included in the catalog for each object is listed and described in Table \ref{tab:header}.

The catalog contains 4 types of flags. The first flag is the modified GALFIT flag (\texttt{flag}) from \cite{vanderWel12}, described in Section \ref{sec:galfit}. We also include a general use flag (\texttt{use}) with values of 1 for ``good'' and 0 for ``bad''. An object is given a flag of 1 if it has a GALFIT flag $\leq1$ (i.e. ``good'' or ``suspicious'' fits) and the effective radius is greater than the PSF FWHM. Otherwise it is flagged with a 0. The third type of flag is a ``parameter robustness limit'' flag (\texttt{flag\_limit\_r}, \texttt{flag\_limit\_n}). There are two of these flags, one for radius and one for S\'ersic index. A galaxy is given a radius robustness flag of 0 if its GALFIT magnitude exceeds (i.e. is fainter than) the radius magnitude limit inferred by its $5\sigma$ depth using Eq. \ref{eqn:rlim} (see Fig. \ref{fig:param_lim}). Similarly, a S\'ersic index flag of 0 is assigned by comparing the GALFIT magnitude to the limit from Eq. \ref{eqn:nlim}. Otherwise, these flags are 1. These flags indicates whether or not an object's effective radius or S\'ersic index can be robustly measured given its magnitude and $5\sigma$ depth. The last type of flag is the deblending flag (\texttt{flag\_deb}), described in Section \ref{sec:sample}. 

A standard selection is obtainable with \texttt{use}=1, though more stringent cuts on the GALFIT flag (\texttt{flag}=0), SNR cuts, and use of the parameter robustness flags (\texttt{flag\_limit\_r}, \texttt{flag\_limit\_n}=0) may be desirable. It is recommended that only non-blended galaxies (\texttt{flag\_deb}=0) are used when pairing COSMOS-DASH morphologies with rest-frame colors from UVISTA, as corrections for blended galaxy colors are not included in the catalog. Blending is most significant at high redshifts and for the most massive galaxies (e.g. $\log(M/M_\odot)>11$). Blended galaxies may have contaminated colors, stellar masses, and/or photometric redshifts, as these quantities are measured from the ground. We recommend that any sample using high mass galaxies, especially at high redshifts, exclude galaxies that are blended.

\section{Quantifying the Utility of DASH Imaging}\label{sec:res}
Owing to the moderately shallow depth of DASH imaging relative to traditional HST imaging, we expect the structural fidelity to diminish for fainter and/or lower stellar mass objects. To probe the parameter space where COSMOS-DASH morphologies are robust, as well as the overall utility of DASH observations, we make direct comparisons to a combined sample of morphologies from other studies. This sample is made up of independent size measurements of galaxies in both the COSMOS-DASH and CANDELS imaging. The bulk of this sample is comprised of a mass-limited galaxy sample selected from all five fields of the CANDELS survey \citep{Koekemoer11}, adopting measurements from the 3D-HST morphological catalogs \citep[hereafter referred to as vdW14]{vanderWel14}. This is augmented by morphologies from COSMOS-DASH imaging for ultra-massive galaxies\footnote{This high mass sample is available at \url{https://archive.stsci.edu/hlsp/cosmos-dash}} \citep[$\log(M_\star/M_\odot)>11.3$;][hereafter referred to as M19]{Mowla19b}. In the M19 sample, COSMOS-DASH morphologies are only measured at $z>1.5$, resulting in a sample size of 203 galaxies. 18 of the galaxies in this sample are split into pairs, and 14 of those remain above the $\log(M_\star/M_\odot)>11.3$ cut. Stellar masses, redshifts, and rest-frame $U-V$ and $V-J$ colors are taken from the UVISTA and are assigned to COSMOS-DASH galaxies based on matching UVISTA ID. In order to maintain consistency between the stellar masses from 3D-HST and the sample from \cite{Mowla19b}, we adopt a $+0.1\,$ dex correction to the UVISTA masses \citep[see appendix B.1 in][]{Mowla19b}. However, further stellar mass corrections done in \cite{Mowla19b} may contribute to differing numbers of high-mass galaxies in the COSMOS-DASH measurements of M19 and the COSMOS-DASH morphological catalog. Moreover, we remove $\sim2\%$ of the ultra-massive COSMOS-DASH galaxies due to deblending flags. 

\begin{figure*}[t!]
    \centering
    \includegraphics[width=0.8\linewidth]{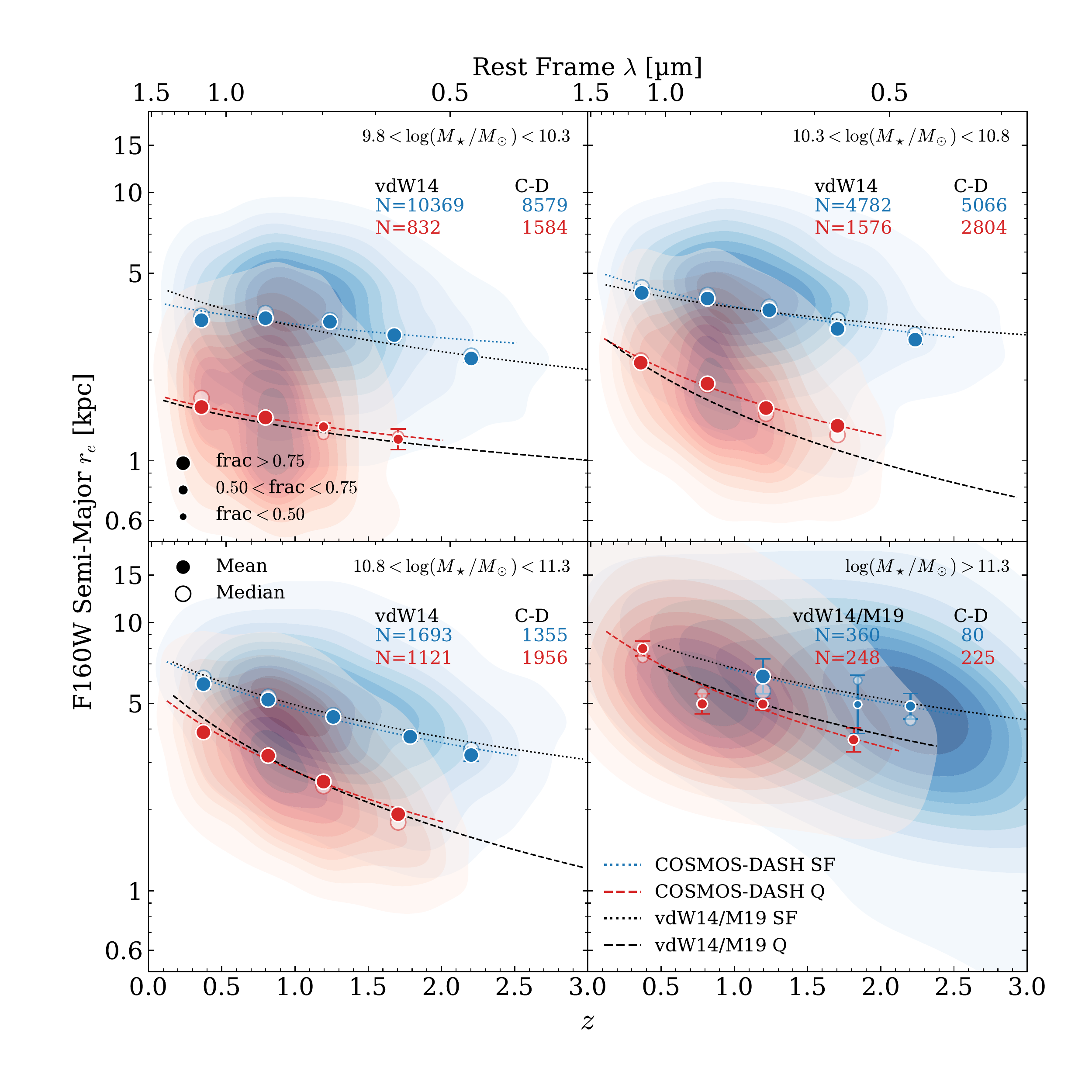}
    \caption{Evolution of quiescent (red) and star-forming (blue) galaxy size with redshift using morphologies from COSMOS-DASH is consistent with size-evolution trends from other studies. Galaxies are separated into mass bins (indicated in the top right of each panel). The distribution of quiescent and star-forming galaxies are shown with colored kernel density estimate contours. The running mean and median for both galaxy types are shown with solid and open points, respectively, and the size of the running mean/median points in each bin varies to reflect the fraction of DASH galaxies in that redshift-mass bin that have a GALFIT flag of 0 or 1 (see Sect. \ref{sec:galfit}). Error bars are determined by the standard error of the mean within each redshift bin. The running means are fit with power laws, see Eq. \ref{eqn:size_evol}, and the corresponding best fit line is shown as a dotted blue or dashed red line for star-forming and quiescent populations, respectively. The same fits are made to the combined sample of galaxies from 3D-HST \citep{vanderWel14} (vdW14) and massive ($\log(M_\star/M_\odot)>11.3$) COSMOS-DASH morphologies \citep{Mowla19b} (M19). These are shown with black dotted and dashed lines for star-forming and quiescent galaxies, respectively. The number of star-forming (in blue) and quiescent (in red) galaxies in each mass bin in the combined sample and COSMOS-DASH (C-D) are listed in the upper right of each panel. Note that vdw14/M19 includes both COSMOS-DASH (at $z>1.5$) and 3D-HST morphologies for the highest mass bin whereas we consider COSMOS-DASH alone. This therefore explains the lower number of galaxies in our analysis.}
    \label{fig:size_evol}
\end{figure*}

\subsection{Size-Evolution}\label{sec:size_evol}
Figure \ref{fig:size_evol} compares the size-redshift evolution relations determined from the COSMOS-DASH morphological catalog to the combined vdW14/M19 sample. Galaxies are separated into star-forming and quiescent classifications using rest-frame $U-V$ and $V-J$ colors. Quiescent galaxies are defined using a modified, redshift-independent version of the \cite{Whitaker12a} rest-frame color separation \citep{vanderWel14,Whitaker15}, given by
\begin{align}\label{eqn:qgal}
    U-V~&>~1.3\text{ \hspace{0.2cm} and \hspace{0.2cm}  }U-V~>~0.8~(V-J)~+~0.7.
\end{align}
We limit the galaxies in the sample to those with GALFIT flags $\leq1$ (see Sec. \ref{sec:galfit}). In Figure \ref{fig:size_evol}, we separate the data into four mass bins and measure the mean (solid points) and median (open points) size evolution of quiescent (red) and star forming (blue) galaxies in. Errors in the mean galaxy size are computed using the standard error of the mean (SEM).

In each redshift bin we also keep track of the ratio of ``good'' or ``suspicious'' fits (\texttt{flag}=0 or 1) to the total number of covered galaxies (\texttt{flag}$\leq3$). This is indicated by the size of each running mean/median point: small points have a fraction $<0.5$, medium points have a fraction between 0.5 and 0.75, and large points have a ratio greater than 0.75. This measurement acts as a proxy for the completeness of the bin. A higher fraction of galaxies with ``usable'' morphologies implies the mean size of that bin better conveys the true average size, while the average size of bins with a lower fraction should be treated with caution. Moreover, we only include redshift bins that have 10 or more galaxies in them, so a sufficient sample size is established. The size evolution with redshift is parameterized with a power law relation \citep{vanderWel14,Mowla19b}, given by
\begin{align}\label{eqn:size_evol}
    r_e=B_z~\times~(1+z)^{-\beta_z}.
\end{align}
This power law is fit to the running mean quiescent and star-forming galaxy size and is shown as the dashed red and dotted blue lines in Figure \ref{fig:size_evol}, respectively.

We compare these power law fits to similar fits to the vdW14/M19 sample. \cite{vanderWel14} and \cite{Mowla19b} use morphological K-corrections to determine ``rest-frame'' sizes, utilizing F125W and F814W below $z=1.5$, respectively, and F160W above \citep[see Appendix B.2 in][]{Mowla19b}. Our COSMOS-DASH morphological catalog only contains F160W sizes\footnote{Though F814W data exists in the COSMOS-DASH area, adding this data is outside of the scope of this particular work}, which trace rest-frame wavelengths redward of the Balmer break for the full redshift range (ranging from 4,000\AA{} at $z=3$ to 16,000\AA{} at $z=0$). To mitigate concerns with K-corrections, we compare observed sizes (without K-corrections) between COSMOS-DASH and vdW14/M19. The vdW14/M19 sample is refit in order to account for these uncorrected sizes. Using rest-frame $U-V$ and $V-J$ colors from the 3D-HST catalogs \citep{Brammer12,Momcheva16} and the UltraVISTA catalog \citep{Muzzin13a}, we separate galaxies into star-forming and quiescent selections. The 3D-HST sample is limited to objects with GALFIT flags $\leq1$ (Sec. \ref{sec:galfit}) from the 3D-HST catalogs. The mean size for each mass and redshift bin is then computed and Eq. \ref{eqn:size_evol} is fit to the running mean size for both quiescent and star-forming galaxies, shown with dashed and dotted black lines in Figure \ref{fig:size_evol}, respectively. The number of galaxies used to measure the size-evolution of quiescent and star-forming galaxies for each mass bin and each data set is also indicated in each panel of Figure \ref{fig:size_evol}.

\begin{table}[t!]
\centering
\caption{Best fit star-forming and quiescent size-evolution relations for both COSMOS-DASH and the vdW14/M19 sample, shown in Figure \ref{fig:size_evol}. Relations are given by Eq. \ref{eqn:size_evol}, as described in Section \ref{sec:size_evol}.}\label{tab:size_evol}
\begin{tabular}{ccccc}
\hline
\hline
\multirow{3}{*}{$\log(M/M_\odot)$} & \multicolumn{4}{c}{\textbf{Star-forming}} \\ \cline{2-5}
 & \multicolumn{2}{c}{COSMOS-DASH} & \multicolumn{2}{c}{vdW14/M19} \\ 
 & $B_z$ & $\beta_z$ & $B_z$ & $\beta_z$ \\ \hline
9.8-10.3 & 3.96$\pm$0.41 & 0.29$\pm$0.14 & 4.60$\pm$0.43 & 0.54$\pm$0.10 \\
10.3-10.8 & 5.22$\pm$0.36 & 0.47$\pm$0.09 & 4.71$\pm$0.35 & 0.34$\pm$0.08 \\
10.8-11.3 & 7.75$\pm$0.40 & 0.71$\pm$0.07 & 7.90$\pm$0.41 & 0.68$\pm$0.06 \\
$>$11.3 & 10.66$\pm$1.40 & 0.68$\pm$0.13 & 10.53$\pm$1.49 & 0.64$\pm$0.14 \\\hline
\multirow{3}{*}{$\log(M/M_\odot)$} & \multicolumn{4}{c}{\textbf{Quiescent}} \\ \cline{2-5}
 & \multicolumn{2}{c}{COSMOS-DASH} & \multicolumn{2}{c}{vdW14/M19} \\ 
 & $B_z$ & $\beta_z$ & $B_z$ & $\beta_z$ \\ \hline
9.8-10.3 & 1.79$\pm$0.03 & 0.37$\pm$0.03 & 1.74$\pm$0.21 & 0.40$\pm$0.19 \\
10.3-10.8 & 3.12$\pm$0.19 & 0.84$\pm$0.09 & 3.22$\pm$0.21 & 1.08$\pm$0.09 \\
10.8-11.3 & 5.78$\pm$0.51 & 1.06$\pm$0.12 & 6.44$\pm$1.02 & 1.21$\pm$0.20 \\
$>$11.3 & 10.42$\pm$1.78 & 1.00$\pm$0.25 & 9.53$\pm$2.00 & 0.83$\pm$0.24 \\\hline
\hline
\end{tabular}
\end{table}

In general, the size evolution measured from COSMOS-DASH is remarkably consistent with that from M19, despite that sample being supplemented with deeper data from 3D-HST/CANDELS and COSMOS ACS. This is especially true for star-forming galaxies (see Table \ref{tab:size_evol}). We note one regime where the average relations deviate by $\sim3\sigma$ in $\beta_z$: intermediate mass quiescent galaxies (i.e. $10.2<\log(M_\star/M_\odot)<10.8$).
Whereas the vdW14 measurements are based on all five deep CANDELS fields in this stellar mass regime, the COSMOS-DASH data covers a significantly shallower but wider area on-sky for the single COSMOS field. The M19 sample uses COSMOS-DASH data in the most massive bin only, whereas we extend the analysis here to significantly lower stellar masses to test the limits of this data. Within this $10.2<\log(M_\star/M_\odot)<10.8$ quiescent sample, which shows a slightly higher average size evolution at $0.8<z<2$, we also find a higher quiescent fraction. The large, contiguous coverage of the COSMOS-DASH survey, as well as known overdensities within the COSMOS field \citep[e.g.,][]{Spitler12,Chiang14} may account for these larger quiescent fractions found in the COSMOS-DASH sample. Overall, despite some limitations, DASH data is sufficient for recovering these relations down to significantly lower stellar mass limits than previously explored when compared to standard HST observations like 3D-HST.

\begin{figure*}[t!]
    \centering
    \includegraphics[width=1\linewidth]{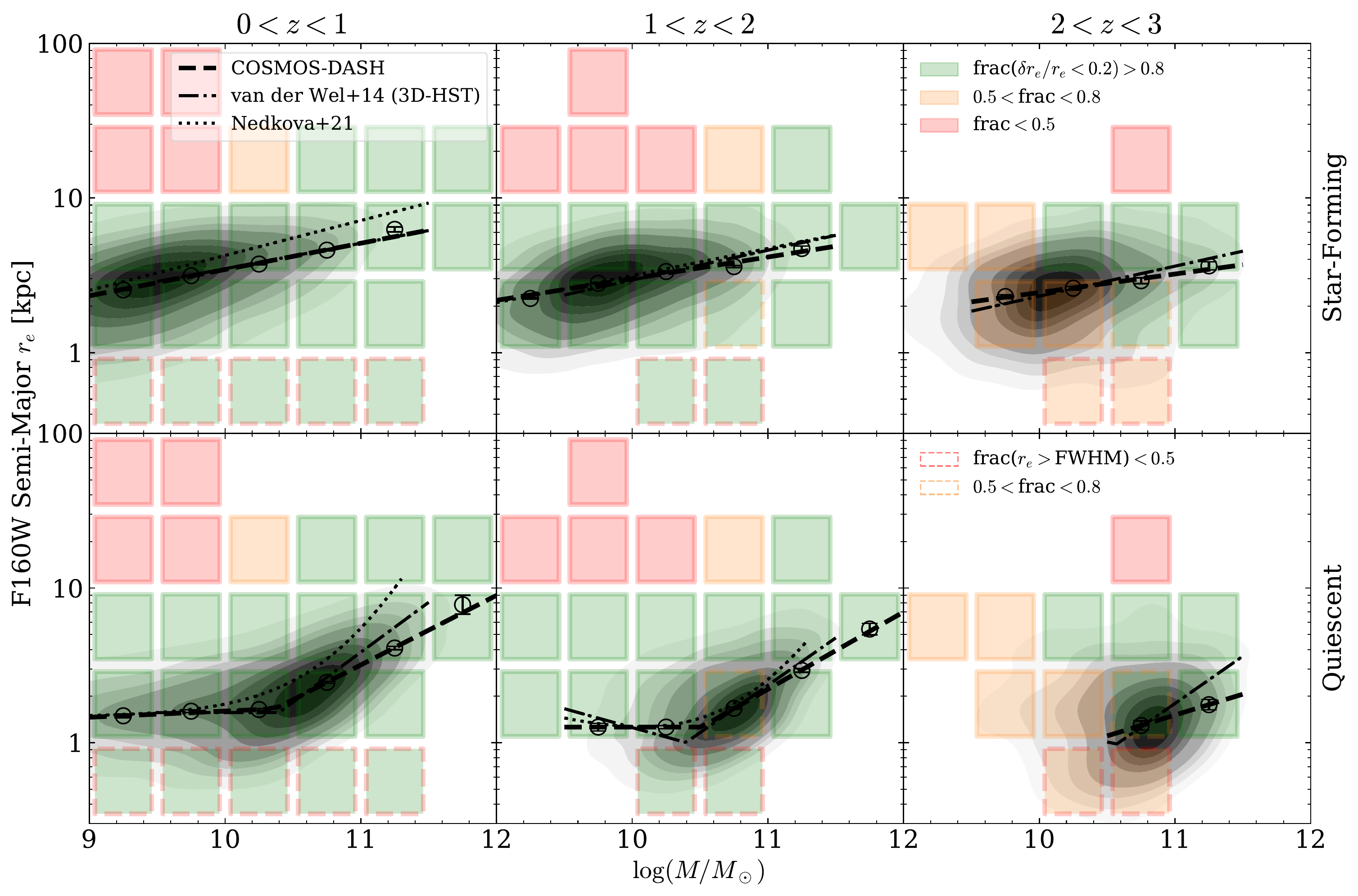}
    \caption{The size-mass relations for star-forming (top row) and quiescent (bottom row) galaxies show COSMOS-DASH sizes are robust for much of the available parameter space. Each column shows galaxies separated into one of three redshift bins: $0<z<1$ (left), $1<z<2$ (center), and $2<z<3$ (right). The underlying distribution of star-forming and quiescent galaxies is shown with contours. Size-mass relations from \cite{Nedkova21} are included for comparison (dotted lines) alongside re-fit size-mass relations (to avoid K-corrections) to 3D-HST morphologies \citep{vanderWel14}. Mass-radius bins are defined by boxes whose colors identify the fraction of galaxies in the bin with fractional size errors less than 20\% ($\delta r_e/r_e<0.2$). Green squares represent a fraction greater than 80\%, yellow show a fraction between 50\% and 80\%, and red indicate a fraction less than 50\%, as shown in the legend in the top right of the right panel. Dashed borders around these boxes indicate a significant fraction of galaxies in that bin have an effective radius less than the PSF FWHM (0.2\arcsec{}). The color of the dashed border conveys the fraction of galaxies in that bin with $r_e>$FWHM: red if the fraction is less than 0.5 and yellow if it is between 0.5 and 0.8.}
    \label{fig:stoplight}
\end{figure*}

\subsection{The Size-Mass Relation}\label{sec:size-mass}
Another useful test of COSMOS-DASH is examining where in size-mass space DASH morphologies are robust via the fraction of galaxies with low uncertainty ($\delta r_e/r_e<20\%$). Figure \ref{fig:stoplight} shows the size-mass relation, limiting our sample to non-blended objects (\texttt{flag\_deb}=0) with GALFIT flags $\leq1$. Using Eq. \ref{eqn:qgal} we classify galaxies in the sample as star-forming and quiescent. Sources are separated into 3 redshift bins and the mean quiescent and star-forming size is measured. Median sizes are only shown in Figure \ref{fig:stoplight} if there are more than 10 objects in the mass bin and $\geq80\%$ of the galaxies in the bin have robust size measurements (\texttt{flag\_limit\_r}=0). Requiring a majority of robust sizes ensures bins with magnitudes fainter than the magnitude limits inferred from Eq. \ref{eqn:rlim} on average are excluded and removes an already known sample of sizes potentially dominated by uncertainties. 

We investigate the robustness of galaxy structural parameters in 0.5 dex size and stellar mass bins, ranging from $\sim0.3$ to 100 kpc and $9\leq\log(M_\star/M_\odot)\leq12$, respectively. For each region in size-stellar mass parameter space with more than 10 objects, the fraction of galaxies with $(\delta r_e/r_e)<0.2$ is computed. Regions are then ranked: fractions greater than 0.8 are ``usable'' and the regions are shaded green in Figure \ref{fig:stoplight}, between 0.5 and 0.8 are labeled ``problematic'' and shaded yellow, and below 0.5 are ``unsuitable'' and shaded red. Some galaxies with $(\delta r_e/r_e)<0.2$ may still not have robust morphologies if the effective radius is smaller than the PSF FWHM (0.2\arcsec{}). To account for PSF limitations, we also compute the fraction of galaxies in each region with $r_e$[\arcsec{}]$<$FWHM. This fraction is reflected in the border of the region. Fractions greater than 0.8 have a solid border the same color as the shaded region, while fractions between 0.5 and 0.8 and less than 0.5 have yellow and red dashed borders, respectively. 

For low redshift galaxies, most of the size-mass parameter space explored is usable, with the exception of very large, low mass galaxies, which are likely nonphysical. These galaxies (as well as intermediate redshift galaxies in a similar parameter space) are likely very faint or noisy, which GALFIT is unable to distinguish from the sky background itself, even after background fitting. The resulting fits attempt to model this, leading to unfeasibly large sizes. For galaxies smaller than 1 kpc the PSF FWHM becomes an issue as well. This is expected, as lower redshift ($z<1$) galaxies, even low mass galaxies, are usually bright and as such are not limited by the depth of DASH observations. At intermediate redshifts $1<z<2$, the usable parameter space is similar to low redshifts, though the FWHM becomes more of an issue owing to the smaller angular sizes. At high redshifts ($z>2$), most of the low mass parameter space becomes problematic and only galaxies with the highest masses yield robust structural parameters. This is likely due to the fact that higher redshift galaxies are fainter (due to distance), causing the shallower DASH depth to become a limiting factor. In general, large, low mass galaxies are unreliable due to GALFIT failing to establish a reasonable fit, while the uncertainty in low mass, high redshift galaxies is due to shallower imaging as a result of the DASH observation technique. Overall, much of the size-mass parameter space remains usable with DASH, with most issues showing up at $z>2$. 

At all redshifts where galaxies are detectable, DASH imaging complements deeper, smaller area surveys. Due to the shape of the mass function, there are fewer massive galaxies in a given area relative to fainter, lower mass galaxies. Since a large sample of lower mass galaxies is obtainable with deeper, narrow field imaging, standard HST observations are suitable, whereas DASH observations may struggle. Likewise, a large sample of massive galaxies requires a larger area survey, though the depth of standard observations is not required, so DASH observations are better suited. This is utilized in \cite{Mowla19b}, where a combined sample of galaxies with a large mass range is assembled using CANDELS/3D-HST morphologies for lower masses ($\log(M_\star/M_\odot)<11.3$) and COSMOS-DASH morphologies for high mass galaxies.

Our analysis of the robustness of structural parameter from COSMOS-DASH suggest that the DASH morphological catalog can be used for lower stellar masses than previously explored in the literature. \cite{Mowla19b} restrict the selection of COSMOS-DASH galaxies to $\log(M_\star/M_\odot)>11.3$ with morphologies from deeper CANDELS/3D-HST catalogs used for lower masses, however our results imply morphologies from COSMOS-DASH, and presumably DASH observations in general, are usable to even $\log(M_\star/M_\odot)\sim9$ for $z<1$.

We also examine the size-mass relation resulting for both quiescent and star-forming galaxies in each redshift range. For star-forming galaxies we parameterize this equation with a single power law, given by
\begin{align}\label{eqn:sfsm}
    r_e~[\text{kpc}]=A~\times~m_\star^\alpha,
\end{align}
where $m_\star\equiv M_\star/7\times10^{10}~M_\odot$ \citep{vanderWel14,Mowla19b}. We then fit Eq. \ref{eqn:sfsm} to the median star-forming sizes and show the best fit lines with dashed lines (Fig. \ref{fig:stoplight}). The best fit parameters for Eq. \ref{eqn:sfsm} are listed in the top of Table \ref{tab:fit}. The behavior of COSMOS-DASH quiescent galaxy sizes at $z<2$ is better described by a broken power law, given by
\begin{align}\label{eqn:qsm}
    r_e~[\text{kpc}]&=
    \begin{cases}
        A~\times~m_\star^{\alpha_1}&M_\star\leq M_p\\
        A\times m_p^{(\alpha_1-\alpha_2)}~\times~m_\star^{\alpha_2}&M_\star>M_p,
    \end{cases}
\end{align}
where $m_\star$ is the same as in Eq. \ref{eqn:sfsm}, $M_p$ is the pivot mass in solar masses, and $m_p\equiv M_p/7\times10^{10}~M_\odot$. We fit this relation to the median quiescent galaxy sizes for the two bins with $z<2$. Due to a lack of low mass galaxies in the high redshift bin ($z>2$) no fit is made. The best fit relations from Eq. \ref{eqn:qsm} are shown with dashed lines in each panel of Figure \ref{fig:stoplight} and the best fit parameters are given in the bottom of Table \ref{tab:fit}.

For comparison, we also fit Eq. \ref{eqn:sfsm} and \ref{eqn:qsm} to the full sample of galaxies from the 3D-HST morphological catalogs \citep{vanderWel14}, again using uncorrected sizes. These relations, as well as the size-mass relations from \cite{Nedkova21}, are shown with dotted and dashed-dotted lines in Figure \ref{fig:stoplight}, respectively. In general, there is good agreement between star-forming size-mass relations at all redshifts. Quiescent size-mass relations are also in agreement at low masses, but the high-mass end shows noticeable deviations. This is discussed in \cite{Mowla19b}, who compare size-mass relations using high-mass COSMOS-DASH galaxies to those from \cite{vanderWel14}. The larger sample of high-mass galaxies obtained with DASH observations may provide a more accurate measurement of the size-mass relation in this mass range, and thus explain the discrepancy. However, it may also to some degree be due to contamination when identifying star-forming and quiescent galaxy populations \citep[as noted in][]{Mowla19b}.

\begin{table}[h!]
\centering
\caption{Results from fits to star-forming and quiescent size-mass distributions given by Eq. \ref{eqn:sfsm} and \ref{eqn:qsm}, as described in Section \ref{sec:size-mass} shown in Figure \ref{fig:stoplight}.}\label{tab:fit}
\begin{tabular}{ccccc}
\hline
\hline
\multirow{2}{*}{z} & \multicolumn{4}{c}{\textbf{Star-forming}} \\ \cline{3-4} 
 &  & $A$ & $\alpha$ & \\ \hline
0-1 & &4.84$\pm$0.08 & 0.17$\pm$0.01 & \\
1-2 & &4.12$\pm$0.15 & 0.15$\pm$0.02 & \\
2-3 & &3.20$\pm$0.09 & 0.12$\pm$0.02 & \\\hline
\multirow{2}{*}{z} & \multicolumn{4}{c}{\textbf{Quiescent}} \\ \cline{2-5} 
 & $A$ & $\alpha_1$ & $\alpha_2$ & $\log(M_p/M_\odot)$ \\ \hline
0-1 & 1.73$\pm$0.04 & 0.04$\pm$0.01 & 0.45 $\pm$ 0.02 & 10.37 $\pm$ 0.03 \\
1-2 & 1.23$\pm$0.03 & -0.01$\pm$0.02 & 0.50 $\pm$ 0.01 & 10.50 $\pm$ 0.01 \\
\hline
\end{tabular}
\end{table}

\begin{figure*}[ht!]
    \centering
    \includegraphics[width=0.8\linewidth]{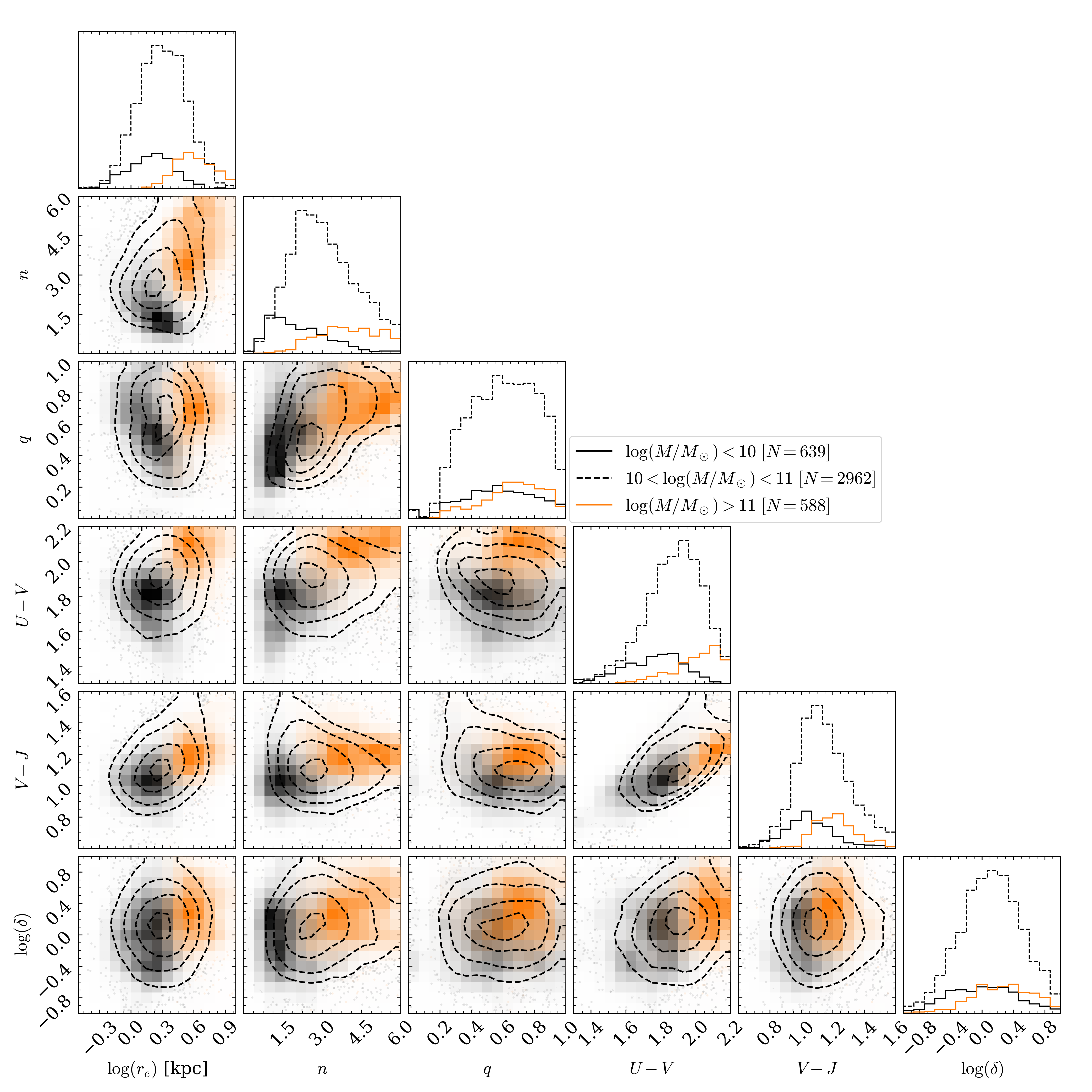}
    \caption{Quiescent galaxies of different mass have different structural properties, as shown in the covariances of various parameters. The included structural and environmental parameters are effective radius, S\'ersic index, axis ratio, $U-V$ and $V-J$ color, and galaxy field overdensity ($\delta=\Sigma/\Sigma_\mathrm{median}$). Low mass ($\log(M_\star/M_\odot)<10$) and high mass ($\log(M_\star/M_\odot)>11$) galaxies are shown with black and orange 1D and 2D histograms, respectively. Dashed contours and 1D histograms indicate intermediate mass galaxies ($10<\log(M_\star/M_\odot)<11$). The number of galaxies in each mass bin is shown in the legend.}
    \label{fig:delts}
\end{figure*}

The result of this broken power law relation is a flattening of the size-mass relation for quiescent galaxies below $\sim3\times10^{10}~M_\odot$. This behavior has been noted previously in the literature \citep[e.g.,][]{Dutton11,Cappellari13,Norris14,Whitaker17,Nedkova21}. Broken power-law relations for all galaxies are also found in \cite{Mowla19a}, with the pivot proposed to mark the transition from star-formation to the dry merger dominated mode of growth for galaxies. This earlier study also compares the size-mass relation to the stellar mass-halo mass relation and finds comparable slopes and pivots between the two relations. Further theoretical support of the flattening of the low-mass quiescent size-mass relation is found in results of simulations \citep{Shankar14,Furlong17,Genel18}. This flattening is likely to be physical and not a result of inaccurate size measurements for small, low mass galaxies \citep[e.g.,][]{vanderWel14, Whitaker17}. We perform a comparative analysis of quiescent galaxy sizes (observed F160W frame) relative to the 3D-HST morphological catalog \citep{vanderWel14}, which yields a similar flattened size-mass relation at low stellar masses. 

\begin{figure*}[ht!]
    \centering
    \subfloat{\includegraphics[width=\linewidth]{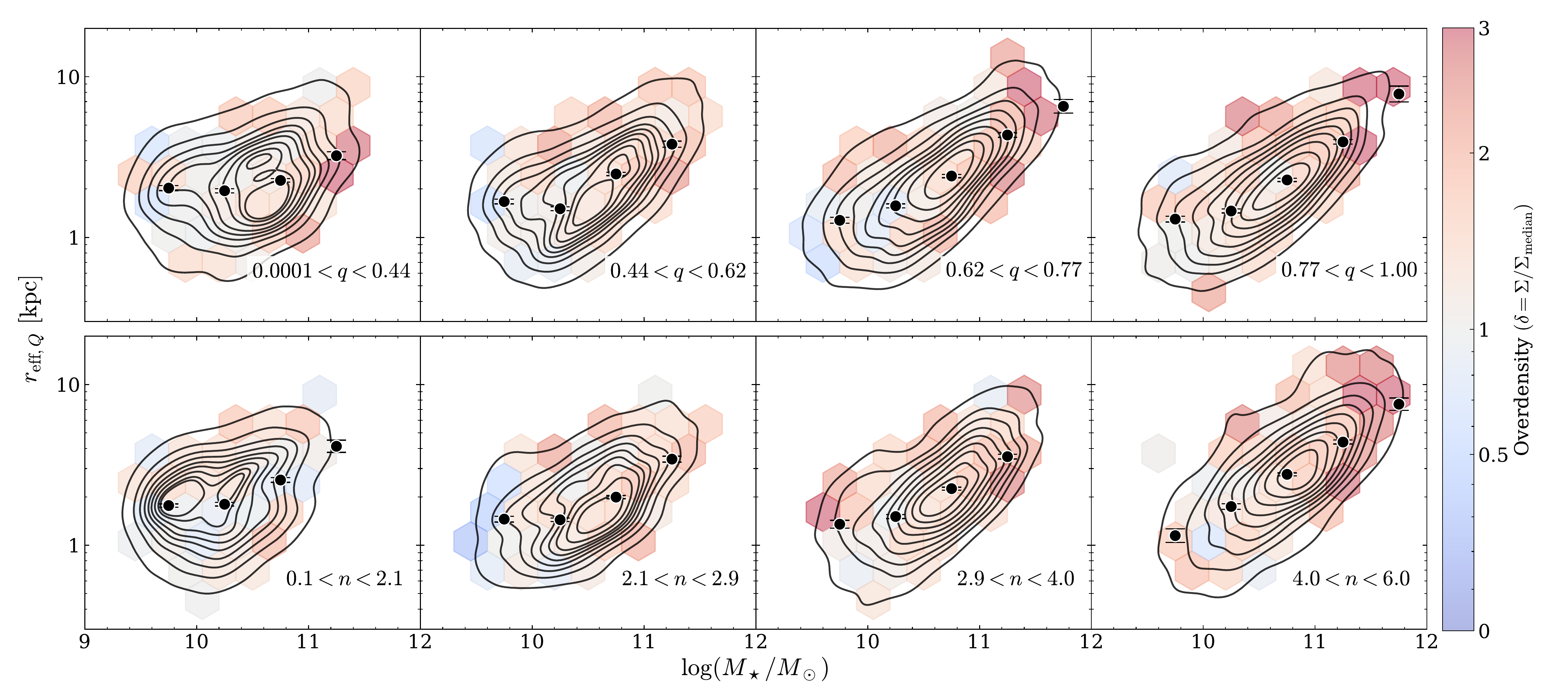}}
    \caption{The flattening of the size-mass relation for quiescent galaxies at $z<1.2$ may disappear in galaxies with high S\'ersic indices. Galaxies are separated by axis ratio (top) and S\'ersic index (bottom) into approximately equal number quartiles indicated in each panel. Empty KDE contours show the underlying distribution of sizes. For each quartile, running medians are shown with solid points and error bars show the $1\sigma$ SEM. Hexagonal bins indicate the average overdensity ($\delta=\Sigma/\Sigma_\mathrm{median}$) of points within the bin, as indicated by the color bar.}
    \label{fig:axisratio}
\end{figure*}

A potential cause of this phenomenon could be the environment in which these low mass galaxies are embedded. There are many ways to form and quench galaxies, and likely different physical processes for cluster and field galaxies \citep[e.g.,][]{PengY10,Tal14,vanDokkum15,Bluck20,McNab21}. Thus, the morphologies of galaxies in different environments could vary. Moreover, as galaxies evolve, their morphologies may change as well \citep{Dressler80,Postman05}. For example, dry mergers can significantly increase the size of quiescent galaxies \citep{Bezanson09,Cappellari13a}, while gas-rich mergers or violent disk instabilities are proposed physical drivers reducing the effective radius of compact star-forming galaxies \citep{Barro13}. We explore the role of environment on quiescent galaxy size in the next section.

\subsection{Environmental Effects}
To investigate the environmental dependence of the flattening of the quiescent size-mass relation, COSMOS-DASH galaxies with $z<1.2$ and GALFIT flags $\leq1$ are matched within a 0.5\arcsec{} radius to the publicly available COSMOS density field catalog \citep{Darvish15,Darvish17}. Since redshift precision is important in determining the galactic density field, we cross match first spatially (using RA and Dec) and then with photometric redshift, discarding galaxies that exhibit a redshift difference of $\Delta z>0.1$ between the two catalogs. This second check removes $\sim5\%$ of galaxies that are cross matched by RA and Dec alone. We then examine the correlation between a number of structural and environmental parameters available for this sample of quiescent galaxies. The parameters used are the size, S\'ersic index, axis ratio, $U-V$ and $V-J$ color, and the overdensity ($\delta=\Sigma/\Sigma_\mathrm{median}$). These correlations are examined for three mass bins: low mass ($\log(M/M_\odot)<10$), intermediate mass ($10<\log(M/M_\odot)<11$), and high mass ($\log(M/M_\odot)>11$), indicated with solid black histograms, dashed black histograms/contours, and orange histograms, respectively, in Figure \ref{fig:delts}.

Figure \ref{fig:delts} highlights different structural parameter distributions (size, S\'ersic index, and axis ratio) for low- and high-mass quiescent galaxies. Lower mass galaxies preferentially have smaller sizes overall, as expected, though these sizes are also similar to intermediate-mass galaxies. This reinforces a flattening of the size-mass relation at low masses. We also find low-mass quiescent galaxies have morphologies similar to star-forming galaxies (smaller S\'ersic index and axis ratio), while their high-mass counterparts are more elliptical in nature (larger S\'ersic index and axis ratio). In general, the distribution of overdensity should be centered around $\delta=1$, since overdensity is defined relative to the median density at a given redshift. However, the distribution of overdensities for low- and high-mass galaxies are skewed, with low-mass quiescent galaxies being more abundant in field ($\delta\lesssim1$) environments and high-mass galaxies being more abundant in cluster ($\delta\gtrsim3$) environments. 

One explanation for the deficit of low mass quiescent galaxies in the highest overdensities could be the preferential destruction of satellite galaxies in clusters via tidal interactions and/or mergers with other cluster galaxies. \cite{Matharu19} found that a significant fraction of compact quiescent cluster galaxies must be destroyed in order to maintain the observed consistency between field and cluster size-mass relations \citep[see also, e.g.,][]{Weinmann09,Maltby10,Cebrian14}. We separate our sample of quiescent galaxies into quartiles of increasing axis ratio (top) and S\'ersic index (bottom) and show the resulting size-mass relations in Figure \ref{fig:axisratio}.

A flattening in the size-mass relation of quiescent galaxies is observed for all S\'ersic index bins except for the most centrally concentrated galaxies. The size-mass relation for quiescent galaxies with $4<n<6$ results because the lowest mass galaxies are slightly more compact, while galaxies close to the pivot mass are slightly larger on average, thereby returning the relation to a power-law. Interestingly, the flattening is still present in the 50\% galaxies with high axis ratios ($q>0.62$). This is driven by face-on disky galaxies, which dominate the low-mass quiescent population: when removing galaxies with $n<2.5$ from the axis ratio analysis, the flattening disappears for $q>0.62$. Quiescent galaxies with lower S\'ersic indices ($n<2.5$) have been measured with a wide range axis ratios, as well as a roughly even distribution over this range \citep[see Fig. 3 in][]{Chang13}. Thus, it is feasible that this sample contains a significant amount of high axis ratio, low S\'ersic index galaxies. These face-on disky galaxies will have comparable axis ratios to centrally concentrated galaxies ($n>2.5$), but are noticeably larger in size. Likewise, when galaxies with $n>2.5$ are removed, the flattening becomes significantly stronger. With the mean overdensity of the galaxies shown within hexagonal bins, we also find that these higher S\'ersic index galaxies inhabit denser environments on average. We note that separating the sample into finer redshift bins does not change the general results of this analysis.

\begin{figure*}[ht!]
    \centering
    \includegraphics[width=\linewidth]{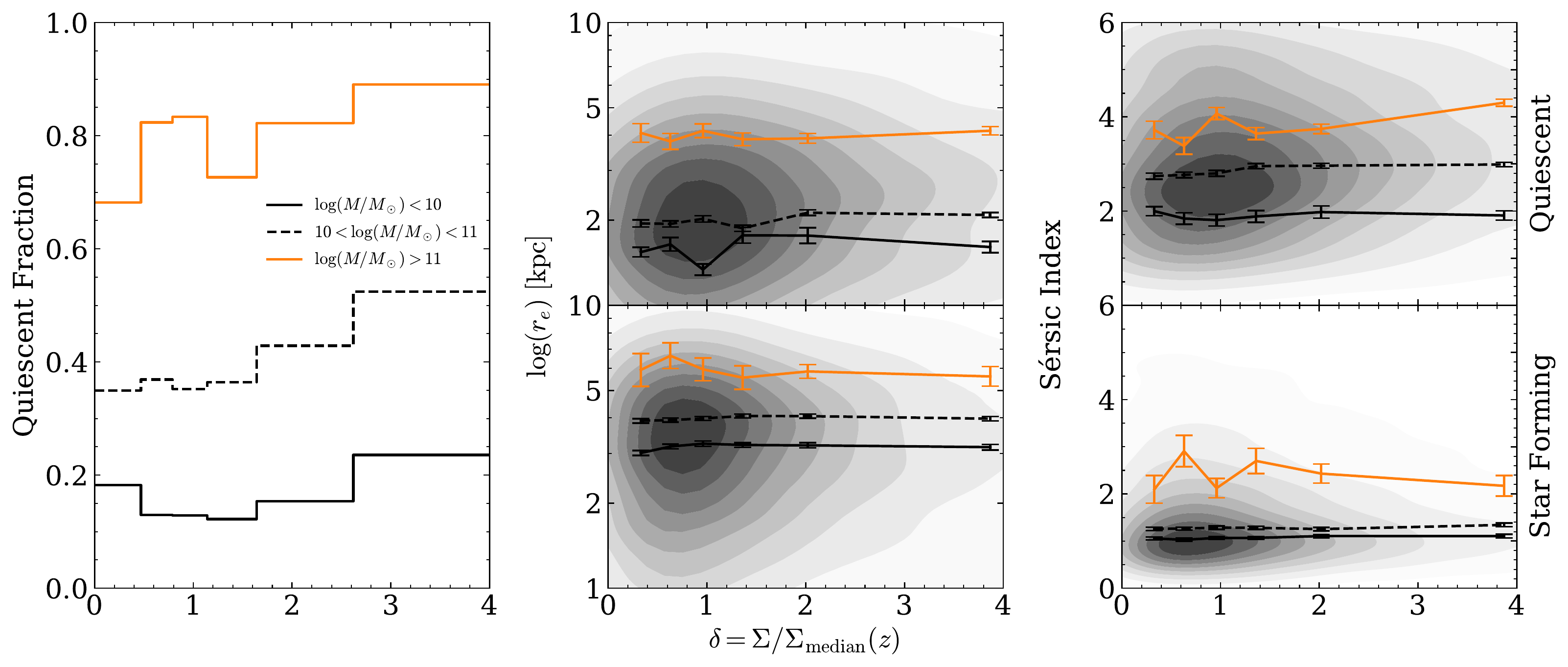}
    \caption{The constant evolution of low ($\log(M/M_\odot)<10$), intermediate ($10<\log(M/M_\odot)<11$), and high ($\log(M/M_\odot)>11$) mass galaxy properties with density suggests environmental effects aren't a dominant mechanism behind the flattening of the quiescent size-mass relation. The three mass regimes are represented with solid black, dashed black, and solid orange lines, respectively. The left panel shows the change in quiescent fraction for each mass bin. The other panels show the change in size (middle column) and S\'ersic index (right column) for both quiescent (top row) and star-forming galaxies (bottom row). The total distributions of quiescent and star-forming sizes and S\'ersic indices are shown with shaded contours.}
    \label{fig:dens_evol}
\end{figure*}

The over-representation of high S\'ersic index galaxies in the densest environments (Figure~\ref{fig:delts}), as well as the existence of compact, centrally concentrated quiescent galaxies in clusters \citep[e.g., right panel in Figure~\ref{fig:axisratio}; see also][]{Matharu19}, may result if disky quiescent galaxies are either preferentially destroyed via mergers or prevented from maintaining a stable disk due to tidal interactions.

Since cluster galaxies do not exhibit a significant increase in size due to minor mergers \citep{Matharu19}, compact cluster galaxies that are not destroyed would retain their smaller sizes. Together, this could explain the disappearance of the flattening at the highest S\'ersic indices; compact elliptical quiescent galaxies are less likely to be destroyed in clusters, and the dynamics of the cluster effectively ``freeze'' their sizes, leading to a single power law quiescent size-mass relation. 

However, the preferential destruction of diskier quiescent galaxies in dense environments should also manifest with increasing S\'ersic index with density. Likewise, if the larger low S\'ersic index galaxies are destroyed in the dense environments, we would expect that low-mass sizes decrease with increasing density. We see no evidence for such a trend in Figure \ref{fig:dens_evol}, which shows the dependence of quiescent fraction (left), size (middle) and S\'ersic index (right) on overdensity. For the lowest mass bin (black solid line), the quiescent fraction remains relatively constant with overdensity, suggesting that the environment does not play a significant role in the quenching of the least massive galaxies. Moreover, both the size and S\'ersic index of similar mass quiescent (top) and star-forming galaxies (bottom) is constant with density for all mass bins. These results imply environment is not the dominant effect producing the flattening of the quiescent size-mass relation at low masses. Though environmental effects may still be significant for some galaxies, universally other processes must drive this behavior. This agrees with the results of \cite{HuertasCompany13}, who find that the size-mass relation of massive ($\log(M_\star/M_\odot)>10.5$) early type galaxies has no significant dependence on environment at $z<0.09$ \citep[see also,][]{Maltby10,FernandezLorenzo13,Shankar14}. 

If environment is not driving the flattening, it may instead be related to in situ physical processes. Studies have shown that the flattening of the size-mass relation is correlated with growth due to star formation, whereas high mass growth is thought to be merger driven \citep[e.g.,][]{vanDokkum15,Mowla19a}. This sets up a narrative where internal processes dominate at lower mass scales whereas extrinsic factors dominate the evolution of the most massive galaxies. The resulting size-mass relation for quiescent galaxies could thus be explained by quenching of low-mass galaxies via the depletion of gas reservoirs, with no replenishment. Another alternative is that this gas is instead stabilized due to the formation of a bulge and thereby inefficient at forming new generations of stars \citep[e.g.,][]{Martig09}, under the assumption that a S\'ersic index of 2 corresponds to a bulge that is sufficiently large. In both instances, the quenching process occurs without mergers, so galaxies remain in a similar location in size-mass parameter space and the shallower slope of the star-forming size-mass relation is retained. While we cannot discern the physical process responsible for this flattening in the quiescent sample within the data set at hand, we can rule out environment as the primary driving factor.  

\section{Summary}\label{sec:sum}
We present a public release of the morphological catalog for the COSMOS-DASH survey obtained with 2D-S\'ersic fits using GALFIT \citep{Peng02,Peng10} for 51,586 galaxies. Using a combination of bootstrapping with GALFIT and comparing morphologies of galaxy analog populations, we obtain parameter errors consistent with those from the CANDELS/3D-HST morphological catalog \citep{vanderWel12,vanderWel14}. We analyze the various limits and parameter space for which these structural measurements and their errors are robust. The wider area, albeit shallower, imaging attainable with DASH is useful for observing a significant sample of high mass, bright galaxies, which are less common in a given area of the sky. This allows DASH observations to work in tandem with standard HST observations: the deeper, standard imaging can obtain significant samples of low-mass, faint sources while wider area DASH observations obtain a sample of high mass galaxies. 

However, DASH observations are also capable of producing results for a wide range of galaxy masses and sizes, as shown in Figure \ref{fig:stoplight}. Analysis of the limits of COSMOS-DASH morphologies suggests sizes and S\'ersic indices are usable to a limiting magnitude of roughly 23 and 22 ABmag for the shallowest DASH observations, respectively. We find DASH sizes at $z<2$ are usable for $\log(M_\star/M_\odot)>9$ and at higher redshifts for $\log(M_\star/M_\odot)>10.5$. Our analysis also indicates that large, unphysical sizes are mainly due to issues with GALFIT, while unusable sizes at higher redshifts are primarily driven by the shallower depth of the COSMOS-DASH survey due to the DASH technique. Overall, the shallower depth and complex data reduction of DASH observations does not prevent high fidelity measurements of galaxy morphology using DASH data.

With the COSMOS-DASH morphological catalog, we observe a flattening of the size-mass relation for quiescent galaxies at low masses ($\log(M_\star/M_\odot)<10.5$), similar to existing results from observations \citep[e.g.,][]{Cappellari13,Norris14,Lange15,Whitaker17,Nedkova21} and simulations \citep[e.g.,][]{Shankar14,Furlong17}. Combining COSMOS-DASH morphologies with galaxy density field measurements \citep{Darvish15,Darvish17}, we investigate the dependence of this low mass flattening on environment. Correlations between structural and environmental parameters suggest distinct morphological populations for low- and high-mass quiescent galaxies, with low-mass galaxies having significantly smaller S\'ersic indices and axis ratios. Moreover, these correlations also indicate that these low-mass quiescent galaxies are more prevalent in underdense environments, relative to their high-mass counterparts. 

Analyzing the evolution of the quiescent size-mass relation with S\'ersic index and environment, we find a disappearance of the low-mass flattening for the highest S\'ersic index galaxies. These centrally concentrated objects are also more likely to inhabit denser environments than their lower S\'ersic index counterparts. Results from other studies \citep[e.g.,][]{Weinmann09,Maltby10,Cebrian14,vanderWel14,Matharu19} imply the majority of satellite galaxies in clusters (dense environments) must be destroyed by $z\sim0$ to match observed size-mass relations. Preferential destruction of compact low S\'ersic index galaxies through tidal interactions or mergers would lead to both an over-representation of compact centrally concentrated galaxies in clusters, as well as a disappearance in the size-mass flattening for high S\'ersic index galaxies. However, the average quiescent fractions, sizes, and S\'ersic indices for both quiescent and star-forming galaxies are all roughly constant with overdensity. Taken together, this suggests that the main driver of the flattening  of the quiescent size-mass relation at low stellar masses is not environment \cite[see also][]{Maltby10,HuertasCompany13,FernandezLorenzo13,Shankar14}. Instead, internal physical processes are likely the cause of the flattening, though exactly which processes are responsible is outside the scope of this work.

\acknowledgments
This paper is based on observations made with the NASA/ESA HST, obtained at the Space Telescope Science Institute, which is operated by the Association of Universities for Research in Astronomy, Inc., under NASA contract NAS 5-26555. These observations are associated with program HST-GO-14114. Support for GO-14114 is gratefully acknowledged. S.C. and K.W. wish to acknowledge funding from the Alfred P. Sloan Foundation grant No. FG-2019-12514, HST-AR-15027, and HST-GO-16259. MA acknowledges support by NASA under award No 80NSSC19K1418. M.S. and K.W. acknowledge support under NASA grant No. 80NSSC20K0416. This research made use of Montage. It is funded by the National Science Foundation under grant No. ACI-1440620 and was previously funded by the National Aeronautics and Space Administration’s Earth Science Technology Office, Computation Technologies Project, under Cooperative Agreement No. NCC5-626 between NASA and the California Institute of Technology. The Cosmic Dawn Center is funded by the Danish National Research Foundation under grant No. 140.

\appendix
\restartappendixnumbering

\section{Background Pedestal}\label{sec:bkg}
\begin{figure*}[t!]
    \centering
    \includegraphics[width=0.8\linewidth]{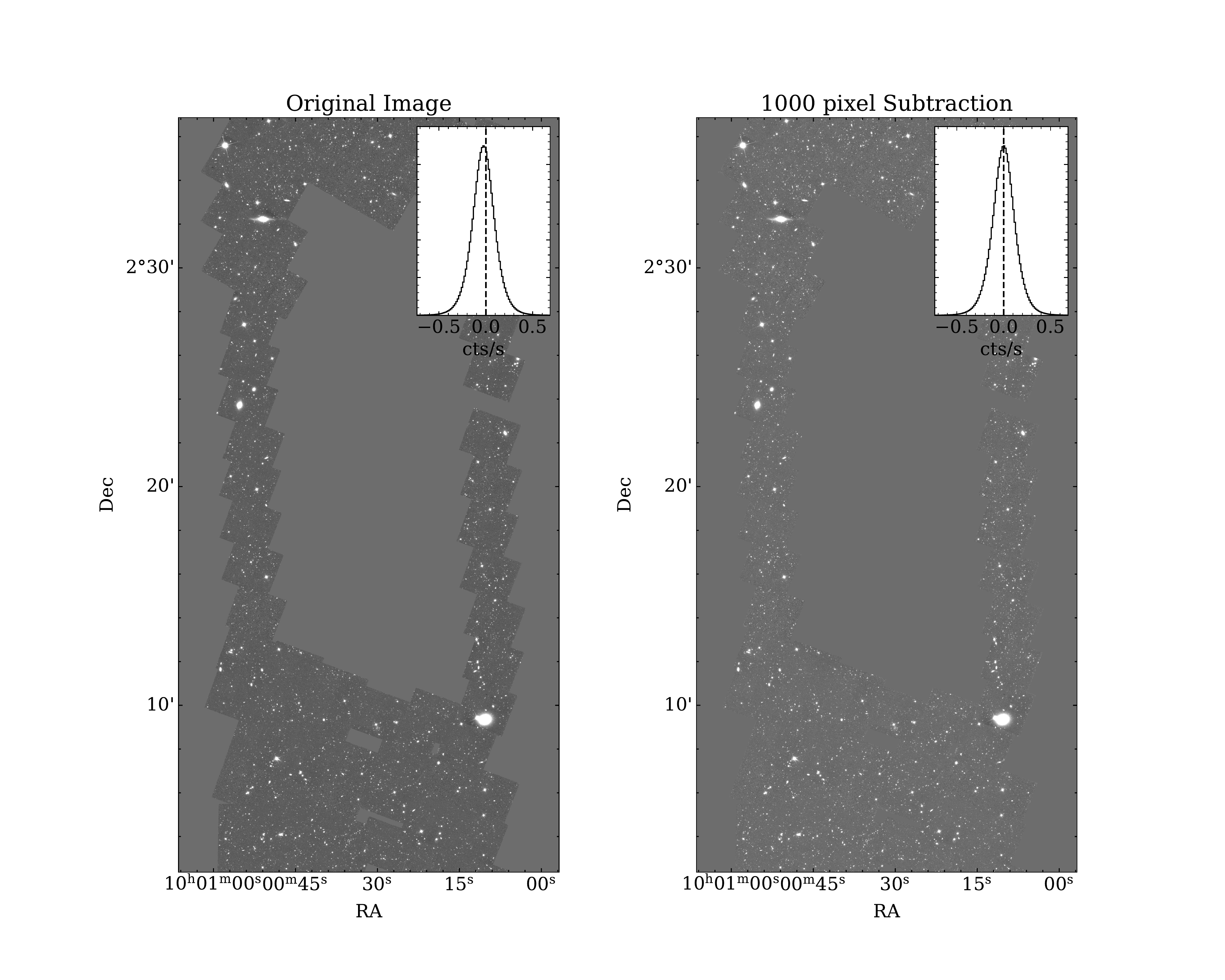}
    \caption{Comparison of the DASH-only reduction of the calibration field before (left) and after (right) background subtraction. The images have the same scale limits, which highlights the slightly negative background in the left panel. The inset histograms show the distribution of background pixels (i.e., any pixel with a value of 0 in the segmentation map). A slight background of $-0.0252$ cts/s is apparent in the histogram in the left panel. This background is much smaller than the standard deviation in background pixels ($0.1465$ cts/s) and is successfully subtracted in the right inset histogram.}
    \label{fig:bkg_comp}
\end{figure*}

Previous studies have shown it is important to let GALFIT fit a constant sky background as a free parameter \citep{Haussler07}. In order to understand the effects of allowing GALFIT to fit the background in more detail we conduct a series of tests. First, estimates of the mean background in the total COSMOS-DASH mosaic are made. Sources are excluded using a combination of the segmentation map and sigma clipping. Bad pixels and empty regions are excluded by removing pixels where the weight map is zero. Separate estimates are made for the DASH and CANDELS regions of the mosaic using a second weight map selection. Low weight pixels are attributed to DASH-obtained data and high weight pixels to CANDELS. The estimated mean DASH background is $-0.0252\pm0.1465$, while CANDELS is $0.0011\pm0.0694$, indicating this DASH data has a more significant background than CANDELS and more variation in the background. This is likely due to the shallower depth of DASH observation caused by shorter exposure times. The DASH background also does not vary significantly as a function of the location in the image.

\begin{figure*}[ht!]
    \centering
    \includegraphics[width=\linewidth]{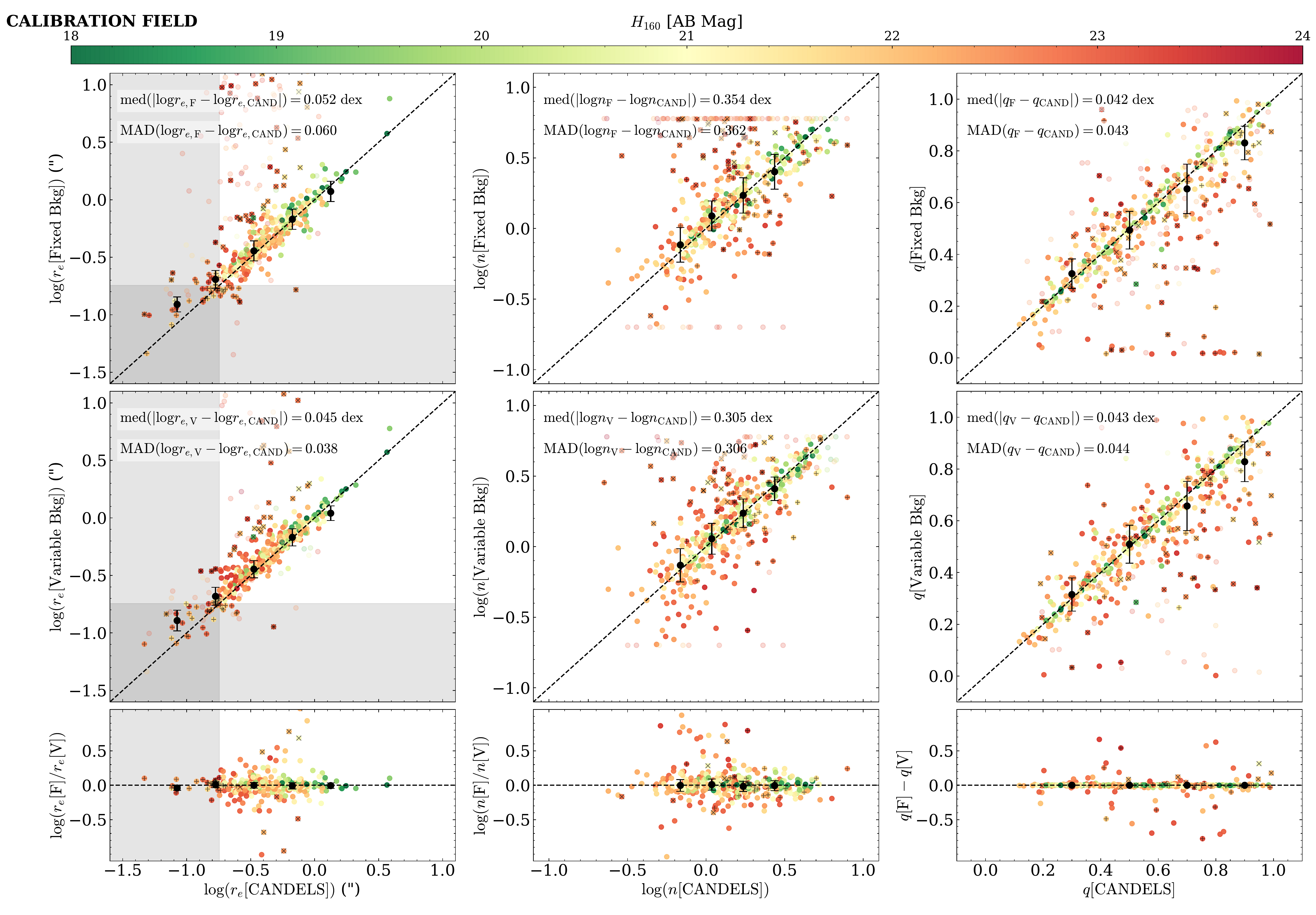}
    \caption{Comparison of the best fit GALFIT parameters of the 483 calibration galaxies in the DASH-only reduction using two different methods to deal with the background; the small scatter in the middle row implies that the background should be treated as a free parameter in GALFIT. The top and middle rows show the effective radii, S\'ersic indices, and axis ratios of the best fit S\'ersic models determined with (top) and without (middle) fixed backgrounds, respectively, compared to those from CANDELS/3DHST \citep{vanderWel14}. Transparent points are those with \texttt{flag}=2 parameter values (see Section \ref{sec:galfit}). The bottom row shows the difference in the parameters when measured both with and without fitting background. The large black points are the running median in each bin, with errorbars showing the scatter determined by the MAD of the bin. The color scale indicates the $H_{160}$ magnitude of the source in AB magnitudes. Points with black x's are those that are greater than 0.3 dex from the one-to-one line in radius. The grey shaded region shows radii less than the PSF FWHM and points with black +'s are sources whose best fit radius falls in this region. Text in the top and middle row indicates the median and scatter in the offset between CANDELS/3D-HST parameters and parameters derived using the two methods to deal with the background.}
    \label{fig:fitvsnofit}
\end{figure*}

We perform a standard background subtraction on both the DASH- and CANDELS-only calibration field reductions. Background meshes with bin sizes of 200, 400, 600, 800, 1000, and 1200 pixels are applied to both reductions and the distribution of background fluxes are computed. The 1000 bin mesh is chosen, as the background-subtracted image for this mesh has a mean background closest to zero. A comparison of the DASH-only reduction before (left) and after (right) background subtraction is shown in Figure \ref{fig:bkg_comp}. First, the GALFIT pipeline is run on the 483 galaxies in the background-subtracted calibration field with the GALFIT sky background fixed to zero. Comparing the resulting best-fit effective radii and S\'ersic indices to the CANDELS/3D-HST morphological catalog \citep{vanderWel12} and high mass galaxies from COSMOS-DASH \citep{Mowla19b}, shows that both the radius and S\'ersic indices are roughly 0.1 and 0.2 dex smaller in DASH, respectively. We then run GALFIT with sky background as a free parameter on the background-subtracted calibration field, finding good agreement with CANDELS/3D-HST. The mean GALFIT-determined sky background is $-0.0206\pm0.0108$, despite measuring a negligible background when the 1000 bin mesh background is subtracted (see also Fig. \ref{fig:empty_aper}, bottom). Similar negligible backgrounds are also measured with the other subtractions. This suggests that this small background value is leading to the significant difference in our measured morphologies.

To further characterize the background of the DASH-only reduction, we use the empty apertures approach from Section \ref{sec:phot} on the background subtracted image. This is done in the DASH-only reduction of the calibration field, which allows for direct comparison with CANDELS (see Fig. \ref{fig:fitvsnofit}). Of interest to this section is the change in the average background as indicated with the shift in the peak of the distribution toward slightly negative values at larger aperture sizes. An increased average background for larger apertures could indicate that measuring the background with a smaller aperture size underestimates the actual background level. However, Figure \ref{fig:empty_aper} (bottom) shows that the average background per pixel is roughly (but not quite) constant with aperture size, and is approximately zero. This behavior is also independent of the overall bin size used to perform the background subtraction. A comparison with the GALFIT background pedestal is also shown, consistent within $2\sigma$. A potential cause of the discrepancy between the GALFIT and empty aperture backgrounds could be that the empty apertures are not empty and contain faint wings from galaxies (potentially due to inadequate masking). 

Next, we investigate whether or not this GALFIT background is something that GALFIT consistently measures from the data, or if GALFIT will only produce reasonable fits if it is allowed to fit a constant sky background. To check this, we run GALFIT twice on calibration field galaxies using the background subtracted DASH-only reduction. In the first run, we allow GALFIT to fit the background. In the second run, we remove the GALFIT-measured constant sky background (see bottom panel of Fig. \ref{fig:empty_aper}) from the data and fix the background parameter to zero. In Figure \ref{fig:fitvsnofit}, the radius, S\'{e}rsic index, and axis ratio of these two fitting methods are compared to each other and the CANDELS/3D-HST Sersic parameters \citep{vanderWel12,vanderWel14}. The top and middle rows of Figure \ref{fig:fitvsnofit} compare the parameters from the run with a fixed and variable (fit by GALFIT) background to the CANDELS/3D-HST morphological parameters, in a similar layout to Figure \ref{fig:seg_comp}. The bottom row compares the difference between the parameters both with and without fixing the background relative to measurements from CANDELS/3D-HST. All panels indicate no systematic offset, both between the different methods to deal with background and between CANDELS/3D-HST and our morphologies. 

In general, both methods show consistency with the measured CANDELS/3D-HST morphologies, suggesting that the small background value measured by GALFIT (see bottom panel of Fig. \ref{fig:empty_aper}) is the root cause of the offset we observed prior. The largest scatter is in S\'ersic index, as this parameter is the most sensitive to the measure of the background. With an incorrect background estimate, the combination of a low S\'ersic index and faint background flux can appear similar to a large S\'ersic index with faint wings. The scatter in the parameters is smaller for radius when we allow GALFIT to fit a background: the fixed background approach has a 9\% larger scatter in radius and a 20\% larger scatter in S\'ersic index. Moreover, 12\% of galaxies with the fixed background are more than 0.3 dex from CANDELS/3D-HST, while only 8\% with a variable background deviate by this much. The median offset from CANDELS/3D-HST (Fig. \ref{fig:fitvsnofit}, text in top and middle rows) also prefers a variable background. This suggests it is preferable to allow GALFIT to deal with the background. Similarly, \cite{Haussler07} conclude that GALFIT performs significantly better when allowed to internally measure a sky background, as opposed to being provided a fixed background (e.g. from SExtractor in their case). Given the smaller spread in morphological parameters and the findings of \cite{Haussler07}, we therefore adopt the background fitting in GALFIT to ensure consistent fits that aren’t biased by the choice of constant background pedestal. Moreover, this analysis also shows that GALFIT consistently recovers morphological parameters when fitting a background regardless of the initial background subtraction. As such, we do not use a background subtracted version of the full COSMOS-DASH mosaic when determining morphologies for the full sample, since GALFIT needs to fit a background regardless of prior subtraction and the background pedestal in the calibration field is roughly constant.

\bibliographystyle{aasjournal}
\bibliography{references}

\end{document}